\documentclass{aa}  
\usepackage[breaklinks,colorlinks,citecolor=blue]{hyperref}
\usepackage{graphicx}

\usepackage{txfonts}
\usepackage{ftnxtra}
\usepackage{color}
\usepackage[flushleft]{threeparttable}

\begin{document}

   \title{The {\tt S-PASS} view of polarized Galactic Synchrotron at 2.3 GHz as a contaminant to CMB observations}

   \subtitle{}

   \author{N. Krachmalnicoff\inst{1}\thanks{e-mail: \href{mailto:nkrach@sissa.it}{nkrach@sissa.it}}
   	\and
	E. Carretti\inst{2,3,4}
          \and
          C. Baccigalupi\inst{1, 6}
          \and
          G. Bernardi\inst{5,7,8}
          \and
          S. Brown\inst{9,10}
          \and
          B.M. Gaensler\inst{11, 12}
          \and
          M. Haverkorn\inst{13}
          \and
          M. Kesteven\inst{3}
          \and
          F. Perrotta\inst{1}
          \and
          S. Poppi\inst{2}
          \and
          L. Staveley-Smith\inst{12,14}
                    }

   \institute{
         SISSA, Via Bonomea 265, 34136, Trieste, Italy
         \and
         INAF Osservatorio Astronomico di Cagliari, Via della Scienza 5, 09047 Selargius (CA), Italy
         \and
          CSIRO Astronomy and Space Science, PO Box 76, Epping, NSW 1710, Australia
           \and
         current address: INAF - Istituto di Radiastronomia, Via Gobetti 101, 40129 Bologna, Italy
         \and
         INAF - Istituto di Radioastronomia, Via Gobetti 101, 40129 Bologna, Italy
          \and
          INFN, Via Valerio 2, 34127, Trieste, Italy
          \and
          Department of Physics \& Electronics, Rhodes University,  Grahamstown, South Africa
          \and
          Square Kilometre Array South Africa (SKA SA), Park Road, Pinelands 7405, South Africa
          \and
          Department of Physics \& Astronomy, The University of Iowa,  Iowa City, Iowa 52245, USA
          \and
          BABL AI Inc., 630 Fairchild St., Iowa City, Iowa 52245, USA
          \and
          Dunlap Institute for Astronomy and Astrophysics, University of Toronto, 50 St. George St, Toronto, ON M5S 3H4, Canada
          \and
          ARC Centre of Excellence for All-sky Astrophysics (CAASTRO)
          \and
          Department of Astrophysics/IMAPP, Radboud University, P.O. Box 9010, NL-6500 GL Nijmegen, The Netherlands
          \and
          International Centre for Radio Astronomy Research, University of Western Australia, Crawley, WA 6009, Australia
          }
\date{}

 \abstract
{We analyze the southern sky emission in linear polarization at 2.3 GHz as observed by the S-band Polarization All Sky Survey ({\tt S-PASS}). Our purpose is to study the properties of the diffuse Galactic polarized synchrotron as a contaminant to $B$-mode observations of the Cosmic Microwave Background (CMB) polarization. \\
We study the angular distribution of the {\tt S-PASS} signal at intermediate and high Galactic latitudes by means of the polarization angular power spectra. The power spectra, computed in the multipole interval $20\le\ell\le 1000$, show a decay of the spectral amplitude as a function of multipole for $\ell\lesssim200$, typical of the diffuse emission. At smaller angular scales, power spectra are dominated by the radio point source radiation. We find that, at low multipoles, spectra can be approximated by a power law $C_{\ell}^{EE,BB}\propto\ell^{\alpha}$, with $\alpha\simeq -3$, and characterized by a $B$-to-$E$ ratio of about 0.5. \\
We measure the polarized synchrotron Spectral Energy Distribution (SED) in harmonic space, by combining {\tt S-PASS} power spectra with low frequency {\tt WMAP} and {\tt Planck} ones, and by fitting their frequency dependence in six multipole bins, in the range $20\leq\ell\leq140$. Results show that the recovered SED, in the frequency range 2.3-33 GHz, is compatible with a power law with $\beta_s=-3.22\pm0.08$, which appears to be constant over the considered multipole range and in the different Galactic cuts.\\
Combining  the {\tt S-PASS} total polarized intensity maps with those coming from {\tt WMAP} and {\tt Planck} we derived a map of the synchrotron spectral index $\beta_s$ at angular resolution of $2^{\circ}$ on about 30\% of the sky. The recovered $\beta_s$ distribution peaks at the value around -3.2. It exibits an angular power spectrum which can be approximated with a power law $C_{\ell}\propto\ell^{\gamma}$ with $\gamma\simeq-2.6$.\\
We also measure a significant spatial correlation between synchrotron and thermal dust signals, as traced by the {\tt Planck} 353 GHz channel. This correlation reaches about 40\% on the larger angular scales, decaying considerably at the degree scales.\\
Finally, we use the {\tt S-PASS} maps to assess the polarized synchrotron contamination to CMB observations of the $B$-modes at higher frequencies. We divide the sky in small patches (with $f_{sky}\simeq1\%$) and find that, at 90 GHz, the minimal contamination, in the cleanest regions of the sky, is at the level of an equivalent tensor-to-scalar ratio $r_{synch}\simeq10^{-3}$. Moreover, by combining {\tt S-PASS} data with {\tt Planck} 353 GHz observations, we recover a map of the minimum level of total polarized foreground contamination to $B$-modes, finding that there is no region of the sky, at any frequency, where this contamination lies below equivalent tenor-to-scalar ratio $r_{FG}\simeq10^{-3}$. This result confirms the importance of observing both high and low frequency foregrounds in CMB $B$-mode measurements.}
 
   \titlerunning{The {\tt S-PASS} view of polarized Galactic Synchrotron at 2.3 GHz}
   \maketitle
 

\section{Introduction}
\label{Section:1}

\begin{figure*}[!t]
\centering
\includegraphics[width=15 cm]{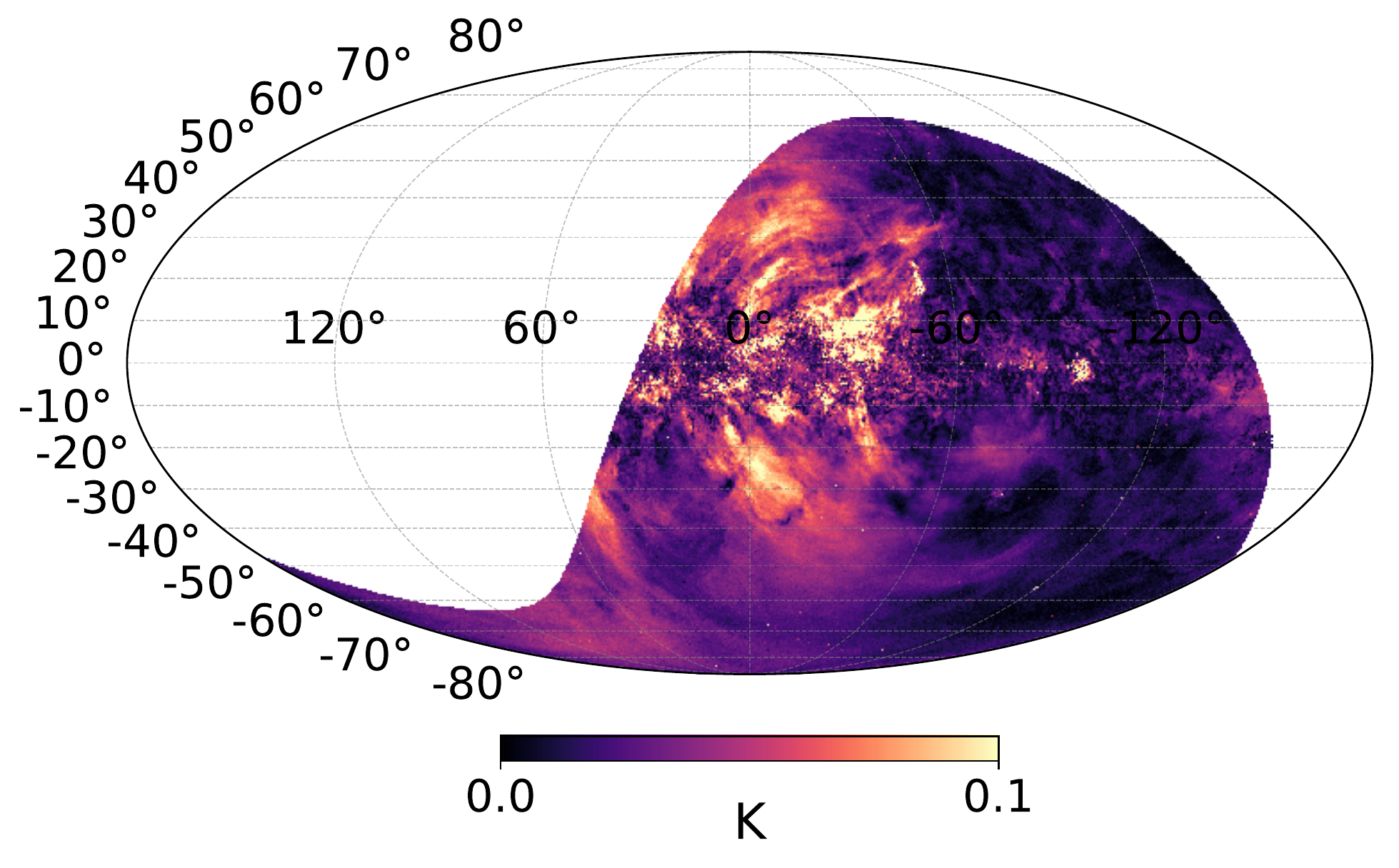}
\caption{{\tt S-PASS} total polarized intensity map  ($P=\sqrt{Q^2+U^2}$) in brightness temperature units and Galactic coordinates.}
\label{SPASS_map}
\end{figure*}

The Cosmic Microwave Background (CMB) anisotropies in linear polarization are sourced by local inhomogeneities at the time of recombination via Thomson scattering \citep[see][ and references therein]{1997NewA....2..323H}. In modern Cosmology, their measurement plays a fundamental role in understanding the physics of the early Universe, the structure formation, as well as the nature of cosmological components.\par

The $Q$ and $U$ Stokes parameters, describing linear polarization, are usually decomposed in $E$ and $B$ modes, representing a gradient and curl component of the polarized field, respectively \citep{1997PhRvD..55.1830Z,1997PhRvD..55.7368K}. The $B$-modes are sourced by non-scalar cosmological perturbations in the early Universe. In particular, primordial Gravitational Waves (GWs) are produced during inflation, an era of accelerated expansion in the very early Universe, responsible for the generation of all cosmological perturbations \citep{2010PhT....63g..49L}. The GWs signature on the polarized CMB signal is characterized by a peak in the angular distribution of the observed $B$-modes power, at degree angular scales. At larger scales, the primordial signal is boosted by a second electromagnetic scattering era, known as re-ionization, causing a bump in the $B$-mode power spectrum at $\ell<20$. On the other side, at smaller scales, the gravitational deflection of CMB photons from forming cosmological structures causes a leak of the $E$ into $B$-modes \citep{1998PhRvD..58b3003Z}. The latter mechanism contributes to the angular distribution of the $B$-mode power, causing a broad lensing peak on angular scales of a few arc-minutes. \par

The CMB anisotropies in total intensity ($T$), $E$-mode polarization, and their correlation $TE$, have been observed by many experiments, culminating with the all sky observations at multiple frequencies, and down to arc-minute angular resolution, by the Wilkinson Microwave Anisotropy Probe \citep[{\tt WMAP}]{2013ApJS..208...20B} and {\tt Planck} satellites \citep{planck2015-I}.\par

Because of the potential breakthrough which would come from the discovery of cosmological GWs, the observation of $B$-modes in CMB experiments has progressed substantially in the last few years. Following early evidences obtained through cross-correlation of lensing, $E$ and $B$ modes by the South Pole Telescope \citep[{\tt SPTpol}]{2013PhRvL.111n1301H}, the lensing peak has been successfully measured by the {\tt PolarBear} experiment \citep{PolarBear14, PolarBear17}, the Atacama Cosmology Telescope \citep [{\tt ACTpol}]{2017JCAP...06..031L}, and others\footnote{See \url{https://lambda.gsfc.nasa.gov/product/expt/} for a list of all operating experiments}.\par

On degree and larger angular scales, the main goal of the experiments is the detection of the inflationary $B$-modes, whose amplitude is usually parametrized through the ratio of the tensor and scalar modes, $r$. Currently, the best upper limit on the tensor-to-scalar ratio is $r<0.07$ at 95\% confidence level obtained from the combination of CMB data with other observational probes (mainly Baryon Acoustic Oscillations, BAO, of the matter density), while the best constraint coming from CMB only observations is $r<0.09$ \citep{2016PhRvL.116c1302B}. The ultimate accuracy of experiments planned for the next decade is expected to reach the level of $r\simeq10^{-3}$, from ground-based facilities such as the {\tt Simons Array} \citep{2017AAS...23030401T}, the {\tt Simons Observatory}\footnote{\url{https://simonsobservatory.org}} and {\tt Stage IV} \citep{2016arXiv161002743A} as well as from space-borne missions \citep{2017arXiv170604516D,2016JLTP..184..824M}. \par

Given the faintness of primordial $B$-modes, a great challenge for the experiments aiming at observing this signal, is represented by the control and removal of diffuse Galactic foreground contamination. At least two mechanisms are active in our own Galaxy generating linear polarized emission: the synchrotron radiation from cosmic ray electrons accelerating around the Galactic magnetic field, and the thermal dust emission from dust grains also aligned with magnetic field. Synchrotron brightness temperature presents a steep decrease in frequency $\nu$ (roughly $\propto\nu^{-3}$) and dominates the sky emission in polarization at frequencies $\lesssim100$ GHz. Thermal dust emission is important at higher frequencies ($\gtrsim100$ GHz) with a frequency scaling well approximated by a modified blackbody with temperature of $\sim20$ K.\par

All sky reconstructions of both components have been obtained from the data of the {\tt Planck} and {\tt WMAP} satellites, covering a frequency range extending from 23 to 353 GHz with sensitivity to linear polarization \citep{planck2015-X}. The analysis of data, targeting explicitly the degree angular scales \citep{planck2014-XXX,K16} indicate that there is no frequency or region in the sky where the foregrounds are proved to be sub-dominant with respect to CMB $B$-modes (at the level of $r\simeq10^{-3}$). Also, analyses that reach the limits of instrumental accuracy show evidence of complex behavior in the Galactic dust and synchrotron emission \citep{planck2017-L, 2017arXiv170909729S, planck2017-LIV, 2015JCAP...12..020C, 2015MNRAS.454L..46Z}. Examples of such potential complexity are represented by the composition and spatial distribution of the dust which, projected along the line of sight, may give rise to de-correlation mechanisms of the polarized emission, i.e. changes in the angular distribution of the signal across frequencies. Line of sight effects, coupled with the spatial distribution of cosmic ray electrons, could in principle affect synchrotron as well. In addition, the energy distribution of cosmic ray electrons could give rise to a curvature in the frequency dependence of the polarized emission, complicating the extrapolation to CMB frequencies. In the near future it might become important to take into account these aspects, implying extra-parameters for estimating, fitting and subtracting the foregrounds in CMB maps.\par 

These considerations motivate two important investigations: on one hand, the development of robust data analysis techniques, capable of removing the foreground component from the data using multi-frequency observations \citep[Component Separation, see ][ and references therein]{2016PhRvD..94h3526S,2016JCAP...03..052E,planck2015-IX}; on the other, the investigation of available data tracing the foreground themselves, in order to determine their complexity. \par

This paper is about the latter investigation. While for thermal dust emission, full sky observations carried out by the  {\tt Planck} satellite at 353 GHz are sensitive enough to allow the study of the main characteristics of the signal and to provide indication of the level of contamination to the current CMB experiments (down to the level of $r\simeq3\times10^{-3}$ at 90 GHz as shown by \citet{planck2014-XXX}), for synchrotron emission the currently available data are less informative. The sensitivity of low frequency observations from  {\tt Planck} and  {\tt WMAP} does not allow the detection of synchrotron signal at intermediate and high Galactic latitudes, where current ground based experiments are observing, and therefore the level of contamination to CMB observation at higher frequency is unclear.\par

To such an aim, radio frequency datasets are being analyzed and will be soon available, specifically at 2.3 GHz from the S-band Polarization All Sky Survey \citep[{\tt S-PASS}]{2010ASPC..438..276C, 2013Natur.493...66C, carretti_etal_in_preparation}, and the C-band All Sky Survey at 5 GHz \citep[{\tt C-BASS}]{2016AAS...22830104P, 2018arXiv180504490J}. These data have a great potential in terms of characterizing the physics of the Galactic medium and magnetic field, but also in terms of studying the synchrotron component as a contaminant to CMB polarization measurements, given the fact that at these frequencies the synchrotron signal is much stronger than at the typical low frequency channels of CMB experiments ($\gtrsim20$ GHz). Moreover, new polarization data will be soon available in the frequency interval 10-40 GHz from the Q and U Joint Experiment at Tenerife \citep[{\tt QUIJOTE}]{2017hsa9.conf...99R} covering about 50\% of the sky in the northern hemishpere. The cross-correlation of all of these data sets ({\tt S-PASS} and {\tt C-BASS} in the south, {\tt C-BASS} and {\tt QUIJOTE} in the north) with {\tt Planck}
 and {\tt WMAP}  will be important to gain insight into the polarized synchrotron behavior, specifically investigating a potential frequency dependence of the spectral index.\par
 
In this paper we describe the analysis of the {\tt S-PASS} data focusing on the study of synchrotron as CMB contaminant; all the survey details are described in a forthcoming paper \citep{carretti_etal_in_preparation}. The most important goals in this analysis are 
related to the study of the angular power spectra of {\tt S-PASS} maps, together with the joint analysis of {\tt S-PASS} data with those at higher frequencies from the {\tt Planck} and {\tt WMAP}, in order to constrain the synchrotron spectral properties, their spatial distribution, and ultimately quantify the level of contamination at CMB frequencies.\par

The paper is organized as follows. In Section \ref{Section:2} we describe the {\tt S-PASS} survey and the data used for the analysis. In Section \ref{Section:3} we characterize the signal by means of the auto angular power spectra on different sky regions. In Section \ref{Section:4} we study the synchrotron frequency dependence using also angular power spectra of low frequency {\tt Planck} and {\tt WMAP} data. A map of the synchrotron spectral index is described in Section \ref{Section:5}. In Section \ref{Section:6} we constrain the spatial correlation between synchrotron and thermal dust emission as observed by {\tt Planck}. In Section \ref{Section:7} we estimate the level of total foreground contamination to CMB $B$-mode signal. Our conclusions are summarized in Section \ref{Section:Conc}.


\section{The {\tt S-PASS} survey at 2.3 GHz}
\label{Section:2}
In this Section we describe the main characteristics of the {\tt S-PASS} polarization maps which have been used in our analysis. For a full description of the {\tt S-PASS} dataset, the observation strategy and the map-making process we refer to \citet{carretti_etal_in_preparation}.

\subsection{The {\tt S-PASS} polarization maps}
The {\tt S-PASS}  survey covers the sky at declination $\delta < -1^{\circ}$ resulting in a  sky fraction $\sim 50$\%. Observations were conducted with the Parkes Radio Telescope, a 64-meter single-dish antenna located in the town of Parkes (NSW, Australia). All details about the survey can be found in \citet{2013Natur.493...66C} and \citet{carretti_etal_in_preparation}, here we briefly report on the main points relevant to this paper. The signal was detected with a digital correlator with 512 frequency channels, 0.5 MHz each. Flux calibration was performed with the sky sources B1934-638 and
B0407-658, used as the primary and secondary flux calibrators, respectively. All frequency channels were calibrated individually, reaching an accuracy of 5\%. The sources 3C 138 and B0043-424 were used as polarization calibrators, for an error on the polarization angle  of $1^{\circ}$. 
Channels with RFI (Radio Frequency Interference) contamination were flagged, and only data in the range 2176-2216 MHz and 2272-2400 MHz were used, binned in one band of 168 MHz effective bandwidth and central frequency of 2303 MHz. The residual instrumental polarization after calibration (leakage of Stokes I into Q and U) is better than 0.05\%.
Total intensity and polarization data are projected on Stokes $I$, $Q$ and $U$ maps, according to the Healpix\footnote{Hierarchical Equal Latitude Pixelization} pixelization scheme \citep{Gorski05}. Final maps have an angular resolution of 8.9 arc-minute and Healpix $N_{side}=1024$ parameter (corresponding to a pixel dimension of about 3.4 arc-minutes).\par

In this analysis we make use only of Stokes $Q$ and $U$ maps\footnote{In this work we follow the COSMO convention for Stokes parameters, which differs from the IAU one for the sign of $U$.}, which completely define the linear polarization amplitude of the observed signal. Figure \ref{SPASS_map} shows the total polarized intensity map obtained from these data ($P=\sqrt{Q^2+U^2}$) in Galactic coordinates and in thermodynamic Kelvin units. The map shows the presence of a diffuse component, due to synchrotron radiation, which dominates the sky polarized emission at this frequency, together with the detection of several point and diffuse radio sources. A bright compact source sample catalogue with polarization information, obtained from {\tt S-PASS} observations, has been presented in  \citet{2016ApJ...829....5L}, while a Stokes $I$ compact source catalogue, complete down to the confusion limit, has been described in \citet{2017PASA...34...13M}. Close to the Galactic plane (for Galactic latitudes $|b|\lesssim20^{\circ}$) the granularity of the emission is indicative of depolarization effects. These are induced by the Faraday rotation of the polarization angle of light propagating through the ionized Galactic medium in the presence of the magnetic field, resulting in different rotations affecting different emission regions across the entire Galactic disc.\par

It is useful to notice that the amplitude of the diffuse synchrotron radiation at 2.3 GHz is a factor of about $10^3$ larger than at 23 GHz (the lowest of {\tt WMAP} frequencies) and about $2\times10^3$ larger than at 30 GHz (the lowest of the {\tt Planck} frequencies). These numbers make it clear the unique benefit of using very low frequency data to characterize the Galactic synchrotron emission, given the high signal-to-noise ratio which can be reached even in those regions where the emission is faint. On the other hand, as we mentioned, observations at low frequencies are affected by Faraday rotation effects, mainly at low latitudes, which can depolarize the synchrotron emission around the Galactic plane, while being marginal at high latitudes. Faraday rotation effects at low Galactic latitude are expected to become milder at frequencies higher than 5-10 GHz.\par

The average noise level on {\tt S-PASS} $Q$ and $U$ maps at $N_{side}=1024$ is about $2.2$ mK, leading to a signal-to-noise ratio larger than 3 on about $94\%$ of the observed pixels. 

\subsection{Sky masks}
\label{Section:mask}
To perform the analysis described in this paper, we use two different sets of masks that we briefly describe in this Section.\par

For characterizing synchrotron emission over large portions of the sky we use a set of six masks obtained as simple iso-latitude cuts in Galactic coordinates. We focus on the sky at intermediate and high Galactic latitudes, where CMB experiments are observing, or plan to observe. In particular, this mask set is composed by cuts at latitudes $|b|$ larger than $20$, $25$, $30$, $35$, $40$ and $50^{\circ}$. We also mask out those regions of the sky where the polarized signal at 2.3 GHz is contaminated by the emission of bright extended radio sources. We mask out the regions around \textit{Centaurus A},  \textit{Large Magellanic Cloud} and \textit{Fornax A}, removing circular patches with radii of $5$, $3$ and $1^{\circ}$ respectively. \par

In order to compute angular power spectra on these sky regions we applied an apodization to the masks, computed with a simple $\cos^2$ function, with apodization length of $3^{\circ}$. The final sky fractions retained for the analysis are of $30, 26, 22, 19, 16, 10\%$ for this set of six iso-latitude masks.\par

The second mask set is composed by smaller sky regions that we use to assess the level of contamination coming from foreground emission to CMB $B$-modes (see Section \ref{Section:7}). In this case we considered 184 regions obtained as circular patches with radius of $15^{\circ}$ centered on the pixel centers of the Healpix grid at $N_{side}=8$ with $|b|>20^{\circ}$. We compute power spectra on these smaller regions also applying a $\cos^2$ apodization to the masks, with length of $5^{\circ}$. The resulting sky fraction for each of the 184 patches in this set of masks is $\sim 1.2\%$. This second set of smaller sky regions is similar to the one used in \citet{planck2014-XXX} and \citet{K16} (the difference being represented by a larger apodization length here for a better power spectrum estimation, given the angular scales of interest), where the level of contamination to CMB $B$-modes coming from thermal dust radiation and synchrotron emission has also been studied.  


\begin{figure*}[!htbp]
\centering
\includegraphics[width=18 cm]{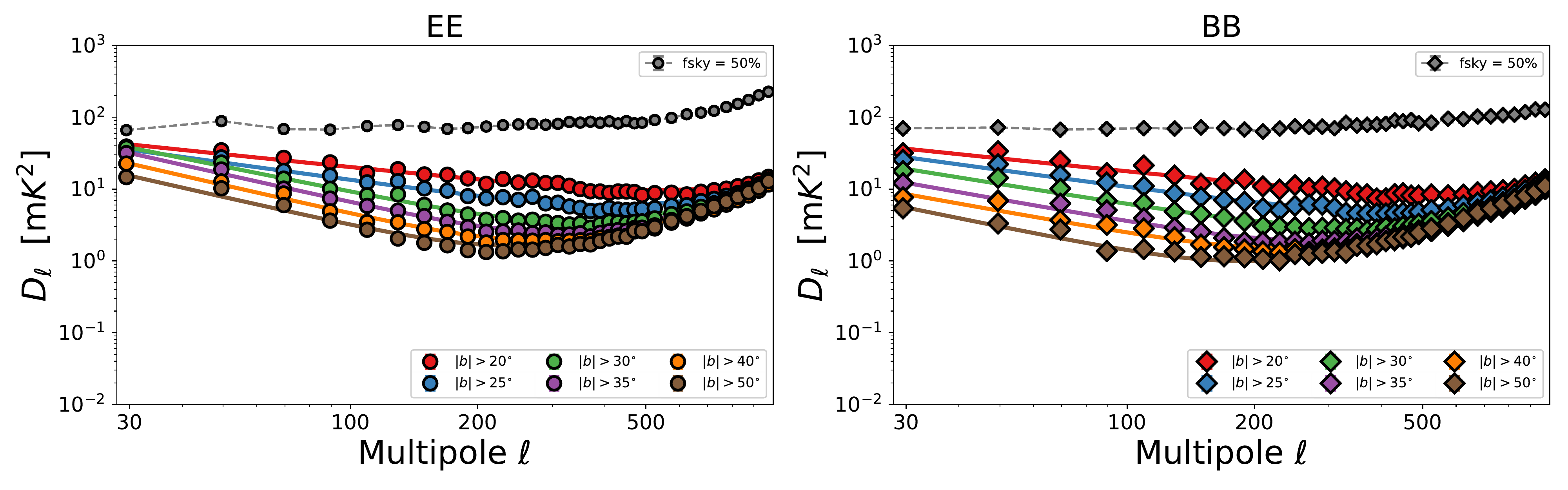}
\caption{$EE$ (left panel) and $BB$ (right panel) power spectra of the {\tt S-PASS} polarization maps computed on the set of iso-latitude masks described in Section \ref{Section:mask} (colored curves). Solid lines show the best fit curve obtained by fitting the model in Equation \ref{spectra_model} to the data (note that in this figure the amplitude of $D_{\ell}=\ell(\ell+1)C_{\ell}/2\pi$ is plotted). Grey curves represent the polarization spectra computed on the whole sky region observed by {\tt S-PASS} where the signal is strongly contaminated by the Faraday rotation effects on the Galactic plane.}
\label{spectra_fit}
\end{figure*}

\begin{figure}[htbp!]
\centering
\includegraphics[width=8 cm]{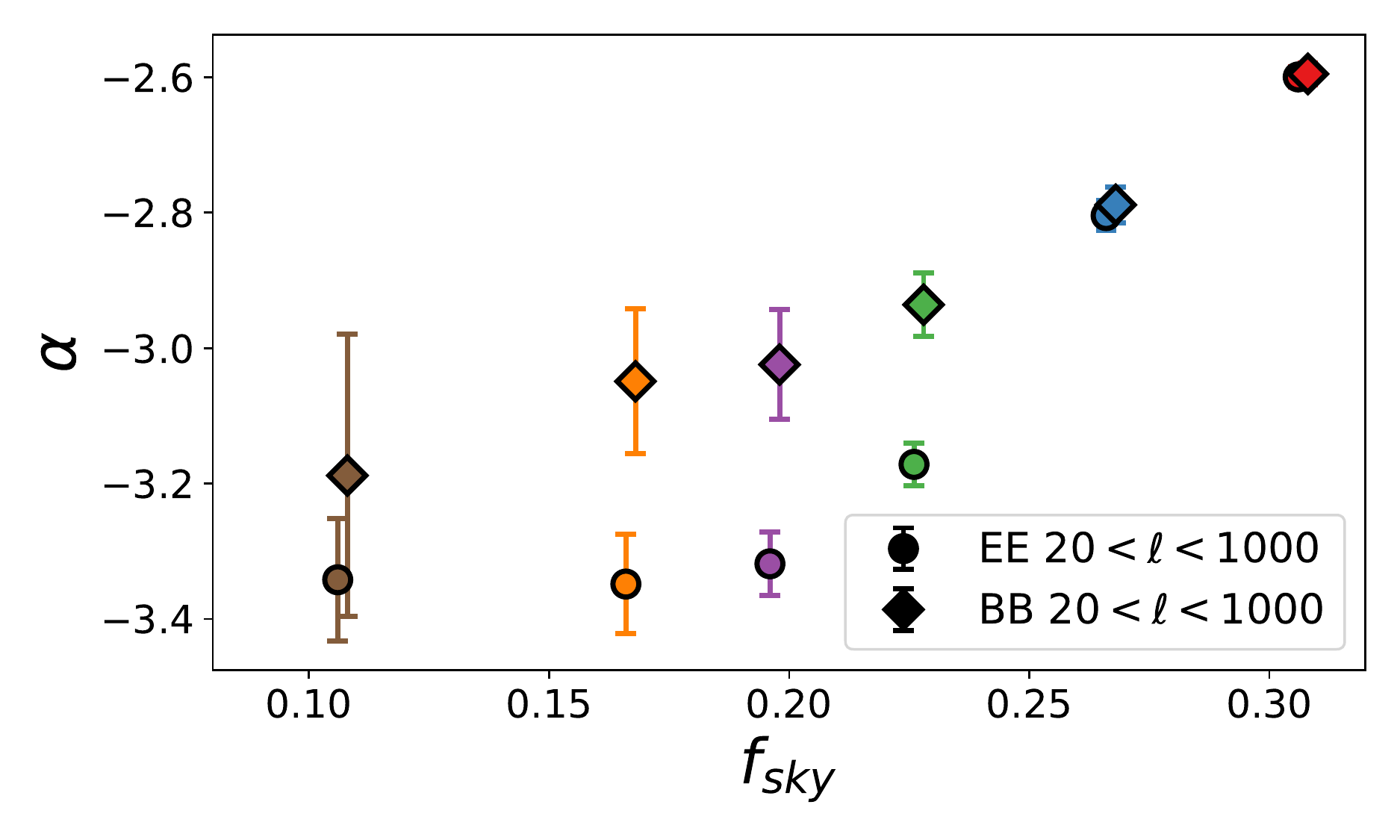}\\
\includegraphics[width=8 cm]{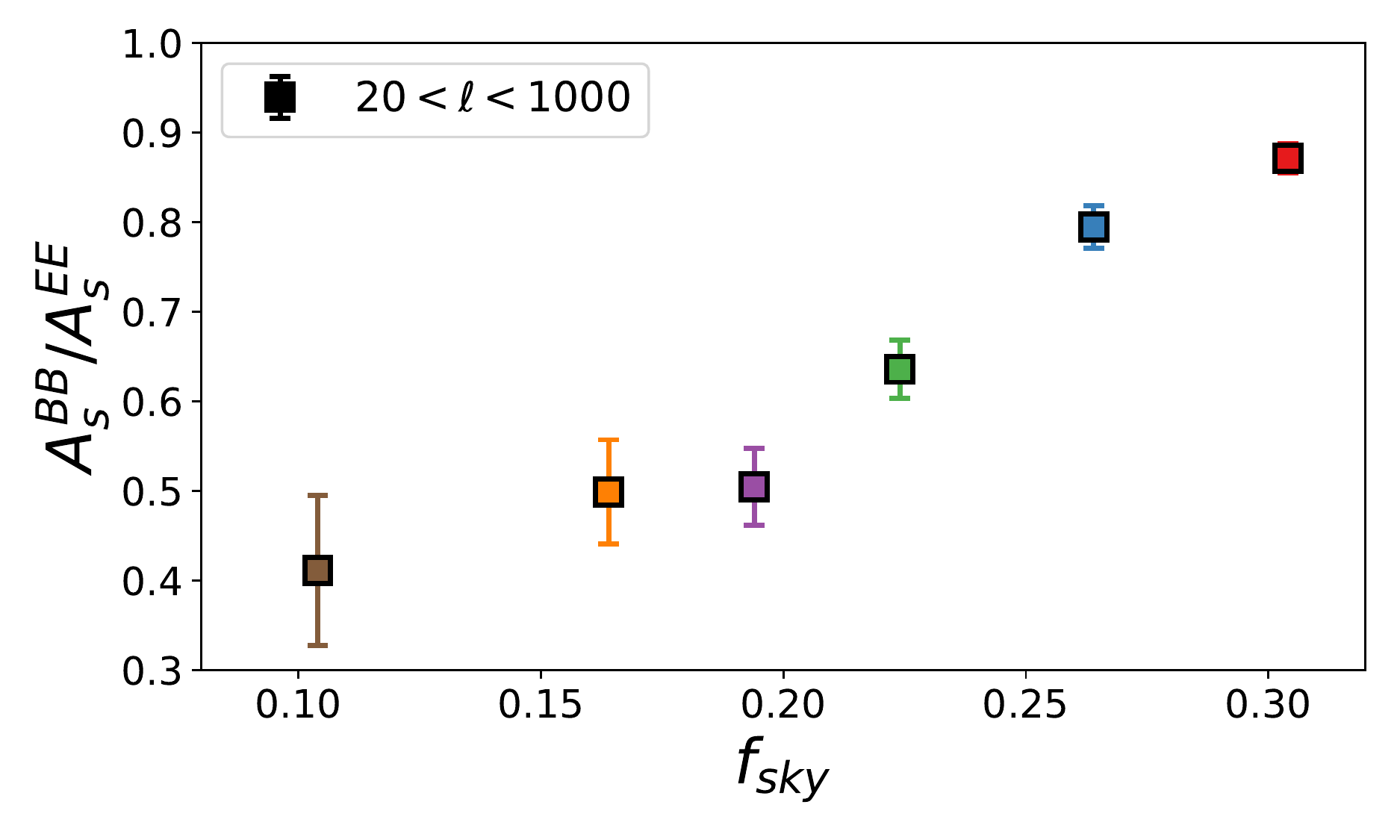}\\
\includegraphics[width=8 cm]{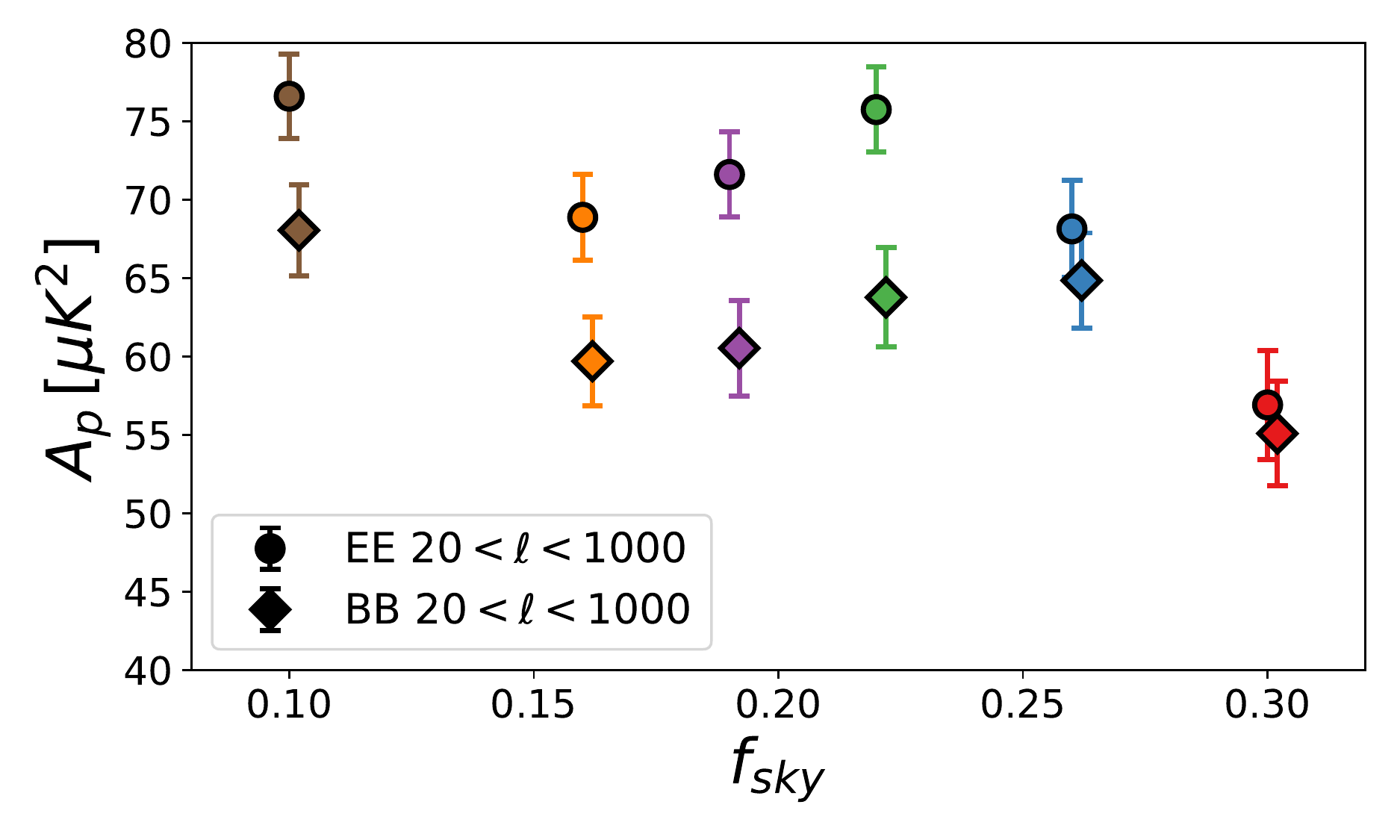}\\
\caption{Best fit parameters obtained by fitting the model in Equation \ref{spectra_model} to {\tt S-PASS} polarization spectra. The color of different points refers to the different sky masks described in the text, following the same color scheme as in Figure \ref{spectra_fit}.}
\label{spectra_fit_results}
\end{figure}

\section{{\tt S-PASS} angular power spectra}
\label{Section:3}

In order to statistically characterize the Galactic synchrotron signal, we compute angular power spectra of the {\tt S-PASS} polarization maps.\par

We compute both $EE$ and $BB$ auto-spectra on the set of six iso-latitude masks, in the multipole interval between 20 and 1000. To correct for the incomplete sky coverage, inducing mixing both between multipoles and polarization states, we compute spectra with an implementation of the {\tt Xpol}\footnote{See \url{https://gitlab.in2p3.fr/tristram/Xpol}.} algorithm \citep{Tristram05}. This code has already been used and tested in several previous works, also for foreground studies \citep[see][and references therein]{planck2014-XXX}. Spectra are binned considering a top-hat band-power with $\Delta\ell=20$.
In the entire multipole range, the noise level lies more than one order of magnitude below the signal power on both $EE$ and $BB$ spectra, and therefore we do not apply any correction for the noise bias.\par

$EE$ and $BB$ angular power spectra for the set of six masks are plotted in Figure \ref{spectra_fit}, showing $D_{\ell}=\ell(\ell+1)C_{\ell}/2\pi$, where $C_{\ell}$ are the two point correlation function coefficients of the expansion in Legendre polynomials. As expected, the amplitude depends on the considered sky region, with spectra computed at higher Galactic latitudes showing less power than the ones including also regions closer to the Galactic plane, where the synchrotron emission is stronger. At low multipoles, corresponding to large angular scales, the sky emission is dominated by the diffuse synchrotron radiation, characterized by a typical power law decay. At $\ell\gtrsim200$, the $D_{\ell}$ start rising due to the emission of extra-Galactic compact sources in the radio band (hereafter, labeled as point or radio sources).

\subsection{Fitting}
\label{Section:spass_specta_fit}

In order to describe the spectral behavior as a function of $\ell$ we implement a fitting procedure considering the following model:
\begin{equation}
\label{spectra_model}
C_{\ell} = A_s^{XX}\left(\frac{\ell}{80}\right)^{\alpha^{XX}} + A_p^{XX}, \,\,\,\,\,\,\,\,\,\,\,\, \text{with }\,XX=EE,\,BB.
\end{equation}
The model characterizes the diffuse Galactic synchrotron emission by means of a power law with index $\alpha$ and amplitude $A_{s}$ evaluated on a pivot point, corresponding to $\ell =80$. This value corresponds to the angular scale where, for CMB emission, the maximum of the contribution from cosmological gravitational waves is located, and therefore most important for measuring the contamination to CMB from synchrotron. The emission from point sources is assumed to be Poisson noise on maps with a flat $C_{\ell}$ power spectrum with amplitude $A_p$.\par

We fitted the model in Equation \ref{spectra_model}, separately for $E$ and $B$-modes and for the different masks, for the whole range of considered multipoles. In performing the fitting we weighted each point with the corresponding signal variance. \par

The fitted power spectra are shown in Figure \ref{spectra_fit} with the resulting best fit parameters reported in Table \ref{fit_results}. Figure \ref{spectra_fit_results} shows the recovered values for $\alpha$ and $A_p$, together with the synchrotron $B$-to-$E$ ratio $A_s^{BB}/A_s^{EE}$. \par

Results show a steep decay of the amplitude as a function of multipoles for the diffuse component of the signal. For iso-latitude cuts with $|b|>30^{\circ}$ the values of $\alpha$ stabilize around $-3.15$ (weighted mean) with a slightly steeper value for $E$ than for $B$-modes. On masks including also low Galactic latitudes the spectrum is flatter, exhibiting a similar behavior for the two polarization states. This is probably due to Faraday rotation of the polarization angle of the signal at mid latitudes, whose effect is to mix $E$ and $B$-modes and to cause excess power on the small angular scales \citep{2001A&A...372....8B}. This consideration is supported also by the behavior of the polarization power spectra computed on the whole sky region observed by {\tt S-PASS} (with the only exclusion of brightest extended sources as described in Section \ref{Section:mask}). As a matter of fact, in this case, where the signal is dominated by  Faraday rotation effects on the Galactic plane, spectra are almost flat in the whole considered multipole range with similar amplitude for $E$ and $B$-modes (grey curves on Figure \ref{spectra_fit}). Given the strong contamination, we do not apply the fit in this case.
\par
The $EE$ spectra show more power than $BB$, especially at high Galactic latitudes, with $A_s^{BB}/A_s^{EE}\simeq0.5$ for $|b|>35^{\circ}$. A similar asymmetry has been observed as a characterizing feature of the thermal dust emission, on the basis of Planck observation at 353 GHz \citep{planck2014-XXX, planck2017-LIV}. A physical explanation of this feature has been proposed for thermal dust emission \citep{planck2016-XXXII} in terms of filaments constituted by dense structures aligned to the Galactic magnetic field. The question whether this reasoning can explain in part or totally the same asymmetry observed here for synchrotron is open. As for the $A_p$ parameter, representing the power of point sources on the spectra, the resulting amplitude is pretty constant for the considered sky regions.\par

It is worth noticing that, in all the considered cases where we fit the model of Equation \ref{spectra_model} to the data, we get high $\chi^2$ values, with Probability to Exceed (PTE) below 5\%. Nevertheless, this is not unexpected: we are modeling the data using a  simple model (although being a typical one for this kind of studies, see for example \citet{planck2014-XXX}), in the high signal-to-noise ratio regime allowed by {\tt S-PASS}. Given the highly non-stationarity and non-Gaussianity of foreground emissions, the spectra are not a perfect power law in $\ell$ and vary in shape and amplitude from region to region. Therefore, the fit may result in a poor correspondence between model and data. Gross statistical indicators, like power law spectra fitting, are important to assess the overall contamination of foregrounds, but are far from constituting a complete knowledge of the signal, which is necessary in order to achieve an effective CMB $B$-mode cleaning.

\begin{table*}
      \caption{Best fit results obtained from fitting the model of Equation \ref{spectra_model} to the {\tt S-PASS} polarization angular power spectra.}
         \label{fit_results}
     $$ 
       \begin{threeparttable}
         \begin{tabular}{lcccccc} 
            \hline\hline
            \noalign{\smallskip}\noalign{\smallskip}
          & $|b|>20^{\circ}$ &   $|b|>25^{\circ}$&  $|b|>30^{\circ}$ &  $|b|>35^{\circ}$ & $|b|>40^{\circ}$ & $|b|>50^{\circ}$ \\           
            \noalign{\smallskip}
             \hline
            \noalign{\smallskip}
              $f_{sky} $            &0.30 & 0.26 & 0.22 & 0.19 & 0.16& 0.10       \\
             \noalign{\smallskip}                                   
            \hline
            \noalign{\smallskip}
                   $A_s^{EE}$ [m$K^2]$ \hfil  & $(2.2\pm0.2)\times10^{-2}$ & $(1.5\pm0.2)\times10^{-2}$ & $(1.1\pm0.1)\times10^{-2}$ & $(8.5\pm0.9)\times10^{-3}$ & $(5.9\pm0.6)\times10^{-3}$ & $(4.0\pm0.4)\times10^{-3}$\\
                    \noalign{\smallskip}
             $A_s^{BB}$ [m$K^2]$ \hfil  & $(1.9\pm0.2)\times10^{-2}$ & $(1.2\pm0.1)\times10^{-2}$ & $(7.3\pm0.8)\times10^{-3}$ & $(4.3\pm0.5)\times10^{-3}$ & $(2.9\pm0.3)\times10^{-3}$& $(1.7\pm0.2)\times10^{-3}$\\
                              \noalign{\smallskip}
             \hline
            \noalign{\smallskip}
            $A_s^{BB}/A_s^{EE}$ \hfil  & $0.87\pm0.02$ & $0.79\pm0.02$ & $0.64\pm0.03$ & $0.50\pm0.04$ & $0.50\pm0.06$ & $0.41\pm0.08$\\
                    \noalign{\smallskip}
             \hline
            \noalign{\smallskip}
            $\alpha^{EE}$ \hfil  & $-2.59\pm0.01$ & $-2.79\pm0.02$ & $-3.16\pm0.03$ & $-3.30\pm0.05$ & $-3.32\pm0.07$ & $-3.31\pm0.10$\\
           $\alpha^{BB}$ \hfil  & $-2.59\pm0.02$ & $-2.78\pm0.03$ & $-2.92\pm0.05$ & $-3.02\pm0.08$ & $-3.03\pm0.12$ & $-3.18\pm0.22$\\
            \noalign{\smallskip}
            \hline
            \noalign{\smallskip}
             $A_p^{EE} [\mu K^2]$ \hfil  & $48.63\pm3.76$ & $61.3\pm3.48$ & $70.82\pm3.56$ & $66.54\pm3.20$ & $64.58\pm3.18$ & $71.85\pm3.35$\\
              \noalign{\smallskip}
             $A_p^{BB} [\mu K^2]$ \hfil  & $49.95\pm4.04$ & $61.19\pm3.82$ & $60.61\pm3.61$ & $57.82\pm3.38$ & $56.89\pm3.38$ & $65.62\pm3.21$ \\
   
             \noalign{\smallskip}
            \hline
         \end{tabular}
     \end{threeparttable}
     $$ 
      \end{table*}
      

\section{Synchrotron spectral energy distribution}
\label{Section:4}
In this Section we describe the results obtained from the analysis of the {\tt S-PASS} polarization data in combination with other publicly available polarization maps, at higher frequencies, coming from the {\tt WMAP} and {\tt Planck} surveys. The goal of this analysis is to characterize the synchrotron Spectral Energy Distribution (SED) on a large interval of frequencies, from 2.3 GHz to 33 GHz. In this Section we summarize the results obtained by analyzing data in the harmonic space, through computation of angular power spectra of the different frequency maps, while in Section \ref{Section:5} we report the results obtained from a pixel-based study.

\subsection{{\tt WMAP} and {\tt Planck} data}
\label{Sec:4.1}
The angular power spectra of the {\tt S-PASS} data and {\tt WMAP}/{\tt Planck} low frequency polarization maps are evaluated using the set of six iso-latitude masks described previously (Section \ref{Section:mask}).\par

We include in our analysis the two lowest frequencies of the {\tt WMAP} dataset, namely the $K$ and $Ka$ bands, centered at  23 and 33 GHz, respectively \citep{2013ApJS..208...20B}. In order to calculate the cross spectra between different frequency maps, we use the full 9-years {\tt WMAP} Stokes $Q$ and $U$ parameters \footnote{\url{https://lambda.gsfc.nasa.gov/product/map/dr5/m_products.cfm}} degraded at the pixel resolution corresponding to $N_{side}=256$. To compute single frequency spectra we cross-correlate the dataset splits, to avoid noise bias. In particular, we split {\tt WMAP} data by co-adding the single year maps from 1 to 4 on one side, and from 5 to 9 on the other. \par

We also include the {\tt Planck} polarizations maps in the 30 GHz band (central frequency 28.4 GHz) obtained from the {\tt Low Frequency Instrument} ({\tt LFI}) observations \citep{planck2015-I, planck2015-II}. Again, for cross frequency spectra we use the full mission maps, while for the case of single frequency we compute the cross correlation of the splits coming from the co-addition of odd and even years maps\footnote{We use the publicly available maps downloaded from the Planck Legacy Archive (PLA) website (\url{http://pla.esac.esa.int/pla}). We use the nominal LFI-30 GHz polarization maps, without applying the correction for the bandpass mismatch. We however checked that our results are not affected by this correction.}.\par

Therefore, our dataset includes four frequencies (2.3, 23, 28.4 and 33 GHz) from which we obtain in total ten power spectra, four computed at a single frequency (corresponding to the one of each single map) and six from the cross correlation among different channels.

\subsubsection{Signal and noise simulations}
\label{Section:sims}
In order to estimate error bars on our set of power spectra we use simulations. In particular, we generate a hundred realizations of signal plus noise maps, as described in the scheme below. 
\begin{itemize}
\item First, we define the simulated polarized synchrotron and CMB radiations included in the simulations. For synchrotron, we consider the sky model adopted in {\tt Planck} Full Focal Plane version 8 (FFP8) simulations \citep{planck2015-XII} that we extrapolate, considering a simple power law SED with $\beta=-3.1$, at the {\tt S-PASS} and {\tt WMAP} $K$ and $Ka$ and {\tt LFI}-30 frequencies. The obtained polarization maps are smoothed with a Gaussian beam for reproducing the angular resolution of the instrument, i.e $FWHM$ (Full Width Half Maximum) of 8.9 arc-minutes for {\tt S-PASS}, 52.8 and 39.6 arc-minutes for {\tt WMAP} $K$ and $Ka$, respectively, and 33.1 arc-minutes for {\tt LFI}-30. Finally we degraded the maps at $N_{side}=256$.
\item We simulate the polarized CMB signal using the \textit{synfast} code of the \textit{healpy} python package. We simulate 100 different maps of the CMB signal, as Gaussian realization of the best {\tt Planck} 2015 $\Lambda$CDM model \citep{planck2015-XIII} considering a tensor-to-scalar ratio $r=0$. Maps are generated at $N_{side}=256$ and convolved with a Gaussian beam corresponding to the angular resolution of each considered frequency channel. 
\item Next, we generate realizations of {\tt WMAP} and {\tt Planck} noise maps, estimating the noise level directly from the data. In particular, we compute the difference maps of the {\tt WMAP} and {\tt Planck} splits described previously. These difference maps, after being properly rescaled with a multiplication factor (1/2 for the case of {\tt LFI} splits which both include half of the total time of {\tt Planck-LFI} observations, and $\sqrt{20/81}$ for {\tt WMAP} splits which include on one side 4/9 and on the other 5/9  of the total time of {\tt WMAP} observations), are a good representation of the noise on the full mission maps. To these maps, we apply the iso-latitude mask with $|b|>20^{\circ}$ and multiply them by the square root of the corresponding full mission map of hits counts, in order to get uniform levels of noise across the sky.\par 
We then compute polarization power spectra of these noise maps and we fit them with a simple $C_{\ell}=A/\ell+B$ model, which, in addition to the flat white noise component, also takes into account the contribution 
$\propto 1/\ell$, coming from the residual $1/f$ noise component on maps. The fit is done in the multipole interval ranging from $\ell=10$ to $200$ which includes the angular scales of interest for this analysis. \par
We generate 100 independent maps of noise at $N_{side}=256$ (for each considered {\tt WMAP} and {\tt Planck} frequency, full mission and split maps) as Gaussian realizations of the obtained fitted spectra. 
We then divide them by the square root of the hits in each pixel, in order to normalize properly the noise level and take into account the sky scanning strategy effect. For {\tt S-PASS} we compute noise maps as random realizations of the variance map (degraded at $N_{side}=256$) which we assume to be the same for both $Q$ and $U$, therefore without including any $1/f$ noise component. The final signal+noise maps are obtained as the sum of the single synchrotron map at each frequency plus the 100 realizations of CMB and noise maps described previously. 
\item Last, once we have our set of simulated maps, we compute cross spectra on them on our six sky masks. We therefore get 100 realizations of our set of 10 spectra. Error bars are then calculated as the standard deviation of these 100 realizations in each considered multiple bin. Since we consider a hundred different realization of the CMB, the final error bars include also cosmic variance. 
\item For {\tt S-PASS} data, we also include in the error budget the calibration uncertainty. In particular we consider a 5\% photometric calibration uncertainty (at the map level) that we add in quadrature to the statistical error. 
\end{itemize}

\subsection{SED fitting}
\label{Section:SED}
\begin{figure*}[!th]
\centering
\includegraphics[width=19 cm]{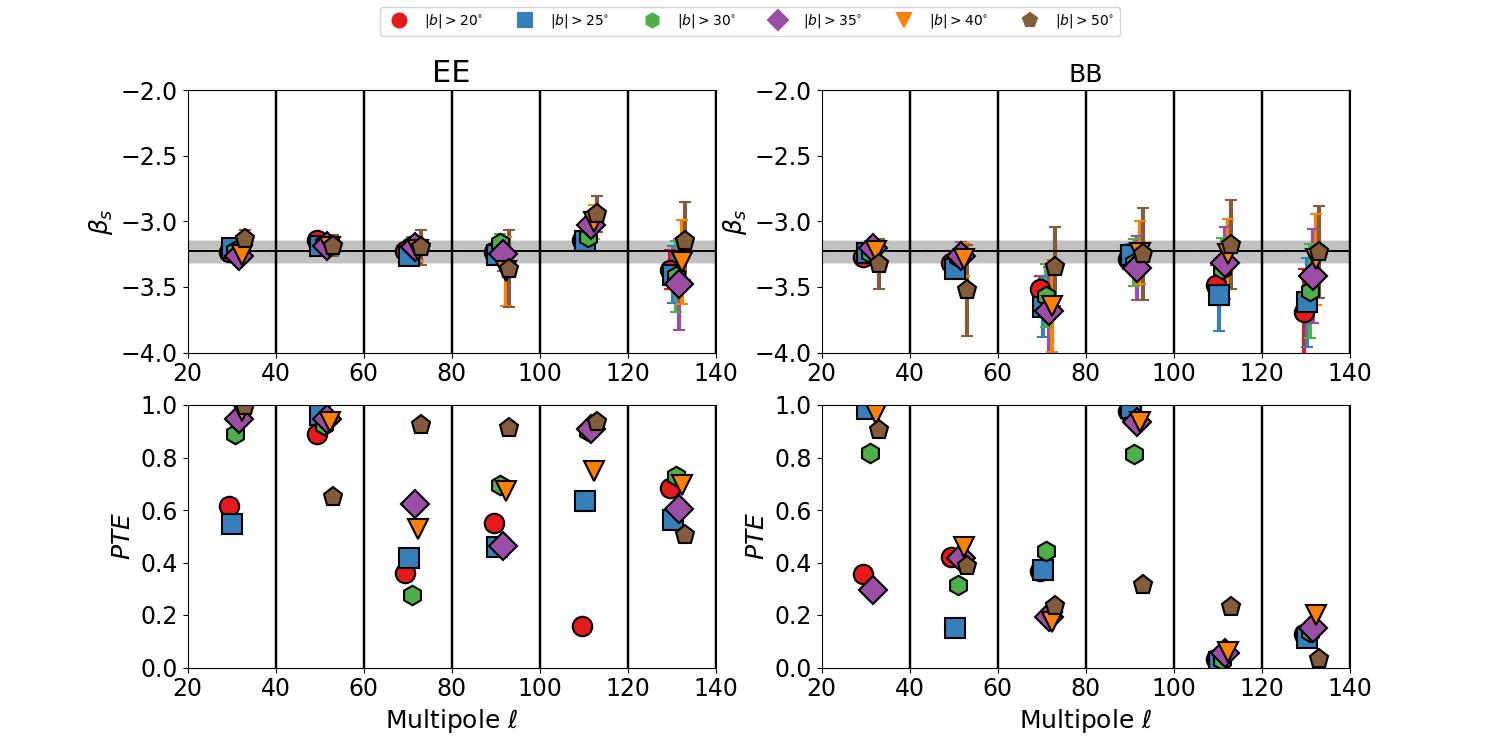}
\caption{Best fit values for the synchrotron SED spectral index $\beta_s$ (upper panel) and PTE coefficients (lower panel), obtained by fitting the model of Equation (\ref{SED1}) to {\tt S-PASS}, {\tt WMAP} and {\tt Planck} data. Different point colors and shapes refer to the different sky regions. The black line and grey area in the upper plot show the retrieved average value $\beta_s=-3.22\pm0.08$.}
\label{fit_nocross}
\end{figure*}

\begin{figure*}[!th]
\centering
\includegraphics[width=16.5 cm]{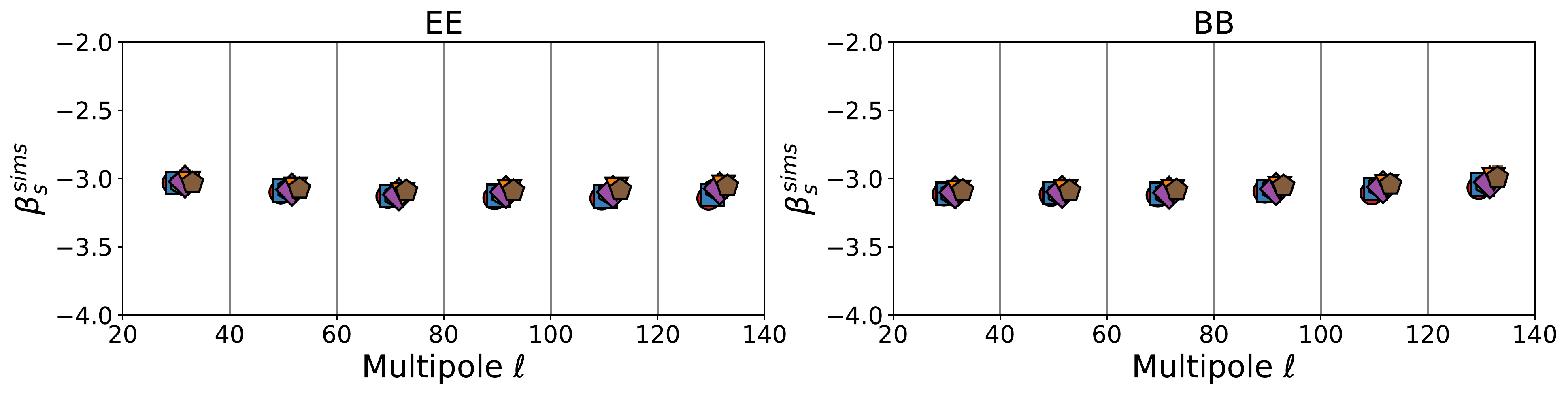}
\caption{Retrieved $\beta_s$ parameter obtained by fitting synchrotron SED model on simulated spectra. Point colors and shapes follows the same scheme of Figure \ref{fit_nocross}. The horizontal line corresponds to  $\beta_s=-3.1$, which represents the input value for our set of simulations.}
\label{fit_sims}
\end{figure*}

\begin{table*}[!t]
      \caption{Best values for the $\beta_s$ parameter and $\chi^2$ obtained from the fit of model in Equation (\ref{SED1}) to data. }
         \label{fit_results}
     $$ 
       \begin{threeparttable}
         \begin{tabular}{lccccccc} 
            \hline\hline
            \noalign{\smallskip}\noalign{\smallskip}
          && $|b|>20^{\circ}$ &   $|b|>25^{\circ}$&  $|b|>30^{\circ}$ &  $|b|>35^{\circ}$ & $|b|>40^{\circ}$ & $|b|>50^{\circ}$ \\           
            \noalign{\smallskip}
             \hline
            \noalign{\smallskip}
              $f_{sky} $ &           &0.30 & 0.26 & 0.22 & 0.19 & 0.16& 0.10       \\
             \noalign{\smallskip}                                   
            \hline
            \noalign{\smallskip}
& $20\leq\ell\leq39$ & -3.23$\pm$0.05 & -3.21$\pm$0.05 & -3.23$\pm$0.05 & -3.27$\pm$0.05 & -3.25$\pm$0.06 & -3.13$\pm$0.07 \cr
& $40\leq\ell\leq59$ & -3.15$\pm$0.05 & -3.19$\pm$0.05 & -3.17$\pm$0.05 & -3.19$\pm$0.05 & -3.2$\pm$0.06 & -3.19$\pm$0.08 \cr
   $\beta_s$ & $60\leq\ell\leq79$ & -3.23$\pm$0.05 & -3.26$\pm$0.06 & -3.18$\pm$0.06 & -3.20$\pm$0.07 & -3.22$\pm$0.09 & -3.20$\pm$0.13 \cr
& $80\leq\ell\leq99$ & -3.23$\pm$0.05 & -3.26$\pm$0.07 & -3.17$\pm$0.07 & -3.25$\pm$0.13 & -3.38$\pm$0.26 & -3.36$\pm$0.29 \cr
& $100\leq\ell\leq119$ & -3.14$\pm$0.06 & -3.15$\pm$0.07 & -3.12$\pm$0.09 & -3.03$\pm$0.09 & -3.01$\pm$0.13 & -2.94$\pm$0.14 \cr
& $120\leq\ell\leq139$ & -3.37$\pm$0.15 & -3.41$\pm$0.22 & -3.42$\pm$0.27 & -3.48$\pm$0.35 & -3.31$\pm$0.32 & -3.15$\pm$0.30 \cr
\noalign{\vskip 3pt\hrule\vskip 5pt}
& $20\leq\ell\leq39$ &          3.6 &          4.0 &          1.7 &          1.2 &          0.8 &          0.3 \cr
& $40\leq\ell\leq59$ &          1.7 &          1.0 &          1.4 &          1.2 &          1.3 &          3.3 \cr
$\chi^2_{N_{d.o.f}=5}$ & $60\leq\ell\leq79$ &          5.5 &          5.0 &          6.3 &          3.5 &          4.1 &          1.4 \cr
& $80\leq\ell\leq99$ &          4.0 &          4.6 &          3.0 &          4.6 &          3.2 &          1.5 \cr
& $100\leq\ell\leq119$ &          7.9 &          3.4 &          1.6 &          1.5 &          2.7 &          1.3 \cr
& $120\leq\ell\leq139$ &          3.1 &          3.9 &          2.8 &          3.6 &          3.0 &          4.3 \cr
 \noalign{\smallskip}
\hline\hline
\noalign{\smallskip}
& $20\leq\ell\leq39$ & -3.27$\pm$0.05 & -3.24$\pm$0.05 & -3.24$\pm$0.05 & -3.21$\pm$0.06 & -3.23$\pm$0.09 & -3.33$\pm$0.19 \cr
& $40\leq\ell\leq59$ & -3.31$\pm$0.05 & -3.37$\pm$0.06 & -3.27$\pm$0.06 & -3.26$\pm$0.10 & -3.29$\pm$0.12 & -3.53$\pm$0.35 \cr
   $\beta_s$ & $60\leq\ell\leq79$ & -3.52$\pm$0.10 & -3.65$\pm$0.23 & -3.57$\pm$0.24 & -3.69$\pm$0.34 & -3.65$\pm$0.35 & -3.35$\pm$0.30 \cr
& $80\leq\ell\leq99$ & -3.29$\pm$0.08 & -3.26$\pm$0.09 & -3.32$\pm$0.18 & -3.36$\pm$0.25 & -3.24$\pm$0.24 & -3.25$\pm$0.35 \cr
& $100\leq\ell\leq119$ & -3.48$\pm$0.15 & -3.56$\pm$0.27 & -3.36$\pm$0.23 & -3.32$\pm$0.27 & -3.25$\pm$0.26 & -3.18$\pm$0.34 \cr
& $120\leq\ell\leq139$ & -3.69$\pm$0.32 & -3.62$\pm$0.34 & -3.53$\pm$0.36 & -3.42$\pm$0.36 & -3.29$\pm$0.34 & -3.23$\pm$0.35 \cr
\noalign{\vskip 3pt\hrule\vskip 5pt}
& $20\leq\ell\leq39$ &          5.5 &          0.7 &          2.2 &          6.1 &          1.0 &          1.6 \cr
& $40\leq\ell\leq59$ &          5.0 &          8.1 &          5.9 &          5.0 &          4.6 &          5.3 \cr
$\chi^2_{N_{d.o.f}=5}$ & $60\leq\ell\leq79$ &          5.4 &          5.4 &          4.8 &          7.4 &          7.7 &          6.8 \cr
& $80\leq\ell\leq99$ &          0.8 &          0.6 &          2.2 &          1.3 &          1.3 &          5.9 \cr
& $100\leq\ell\leq119$ &         12.1 &         13.1 &         12.6 &         10.8 &         10.7 &          6.9 \cr
& $120\leq\ell\leq139$ &          8.5 &          8.9 &          8.4 &          8.1 &          7.3 &         12.2 \cr
\noalign{\vskip 3pt\hrule\vskip 5pt}
         \end{tabular}
     \end{threeparttable}
     $$ 
      \end{table*}

\begin{figure*}[!tp]
\centering 
\vspace{20 pt}
\includegraphics[width=\textwidth]{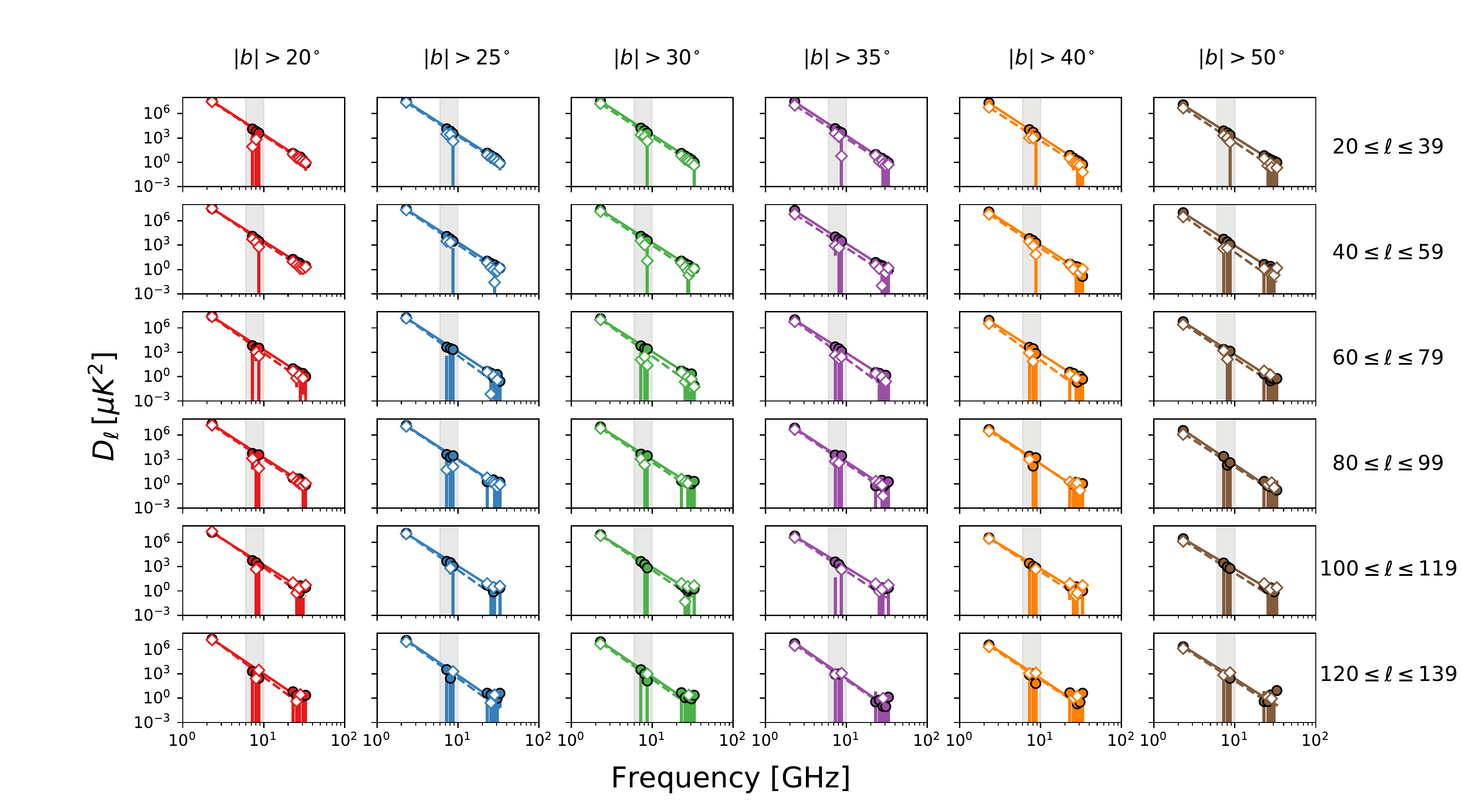}\\
\vspace{20 pt}
\includegraphics[width=\textwidth]{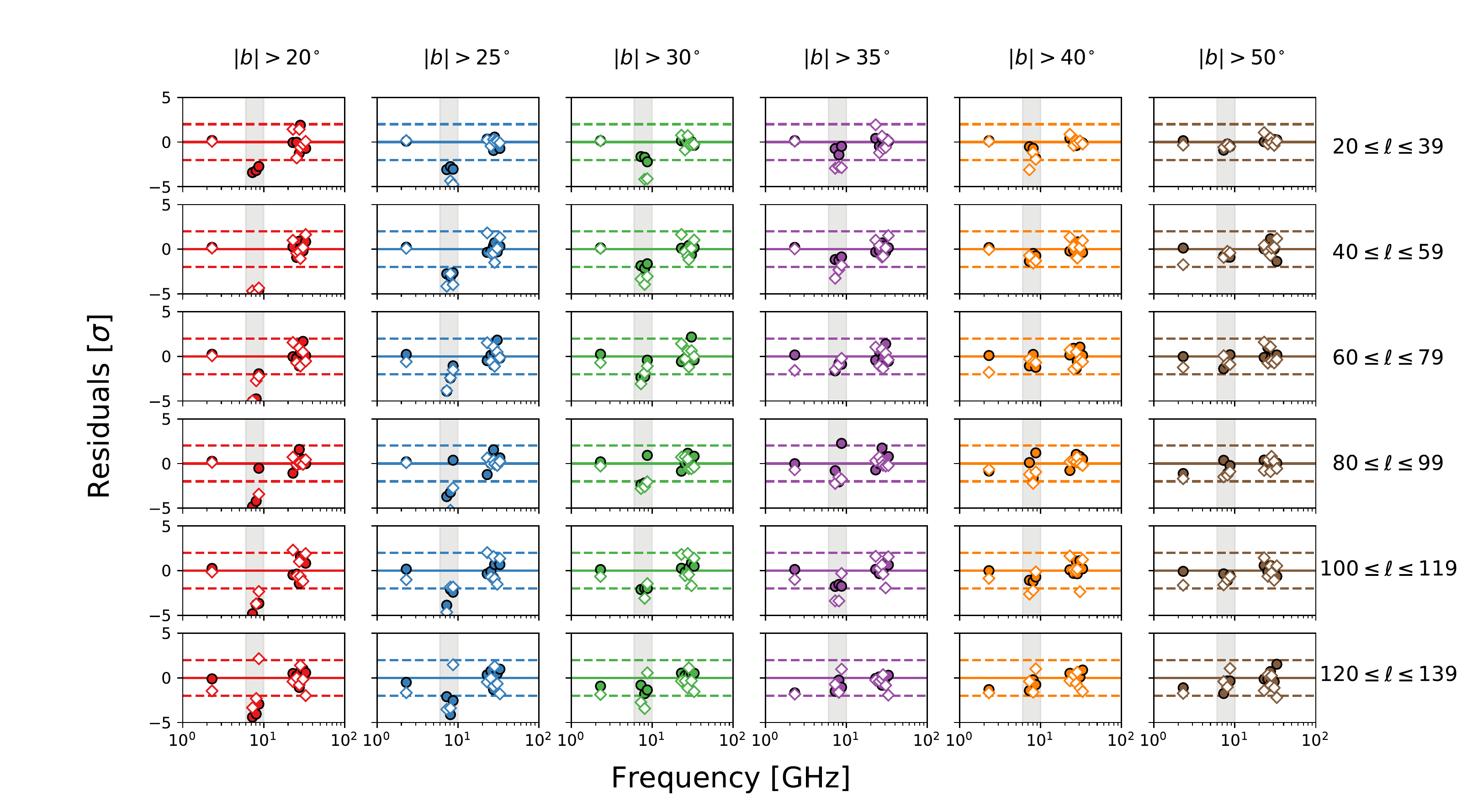}
\caption{Upper panel: amplitude of computed power spectra on data for different multipole bins and sky masks (the color scheme is the same of Figure \ref{fit_nocross}) as a function of the effective frequency. Filled points refers to $EE$ spectra, empty ones to $BB$. Curves represent the best model we obtain when fitting Equation (\ref{SED1}) to data (solid and dashed lines for $E$ and $B$-modes respectively). Note that in this figure the amplitude of $D_{\ell}=\ell(\ell+1)C_{\ell}/2\pi$ is plotted. Lower panel: residuals of the fits normalized to the $1\sigma$ error, the dashed horizontal lines represent the $2\sigma$ level. In both upper and lower panel the points inside the grey shaded area come from the correlation of {\tt S-PASS} with {\tt WMAP/Planck} data which, as described in the text, are not considered in the fitting.}
\label{all_SED}
\end{figure*}

As mentioned, we compute power spectra on our set of six iso-latitude masks, using the {\tt Xpol} power spectrum estimator, on multipoles ranging from $\ell=20$ to $\ell\simeq140$. In particular, we consider top-hat band powers in the following multipole bins: $[20,39]$, $[40, 59]$, $[60, 79]$,  $[80,99]$,  $[100,119]$,  $[120,139]$. We restrict our analysis to the large angular scales, in order to maximize the signal-to-noise ratio on {\tt WMAP} and {\tt Planck} data, and to avoid contamination from point sources in {\tt S-PASS} data. Power spectra are corrected for the beam window function, considering Gaussian beam profiles at the nominal angular resolution of each channel.\par

We use the spectra to fit for the synchrotron SED. In particular we perform the fit separately for each sky mask, for each multipole bin and for $B$ and $E$ modes, ending up in doing 72 different fits (6 sky masks $\times$ 6 multipole bins $\times$ 2 polarization states).\par

We fit the following model to data:  
\begin{equation}
C_{\ell}(\nu_1\times\nu_2) = A_s\,\left(\frac{\nu_1\nu_2}{\nu_0^2}\right)^{\beta_s},
\label{SED1}
\end{equation}
where $C_{\ell}(\nu_1\times\nu_2)$ represents the amplitude of the cross power spectrum between frequency $\nu_1$ and $\nu_2$ (in GHz), in a given multipole bin and mask, and it is expressed in terms of squared brightness temperature units. We fix the pivot frequency at $\nu_0=9$ GHz, close to the geometrical mean of the considered frequency range. \par

At this stage of our analysis we fit this simple model, whose free parameters are the amplitude $A_s$ of synchrotron radiation at the pivot frequency of $9$ GHz and the spectral index $\beta_s$, to a subset of the available frequencies. In particular, we restrict the fit to the {\tt S-PASS} auto spectra, the  {\tt WMAP} and {\tt Planck} single frequency spectra, and all the possible cross spectra between {\tt WMAP} and {\tt Planck}, excluding cross-spectra involving  {\tt S-PASS}. Therefore, we fit our model to data at seven different effective frequencies (where as effective frequency we consider the geometrical mean $\nu_{eff} = \sqrt{\nu_1\nu_2}$), being: 2.3, 23, 25.5, 27.5, 28.4, 30.6 and 33 GHz. We restrict to this subset of frequencies in order to avoid possible biases due to frequency de-correlation \citep{planck2017-L} making the model in Eq. \ref{SED1} not appropriate.  This effect is negligible, due to noise, in cross spectra between {\tt WMAP} and {\tt Planck}, but could play a role in the case of cross spectra involving {\tt S-PASS}. We address the problem of de-correlation in Section \ref{Sec:decorr}. \par

Before performing the fit, we correct our power spectra amplitude by applying the color correction coefficients to {\tt WMAP} and {\tt Planck} data, in order to account for the effect introduced by the finite width of the frequency bands. In particular, we use the coefficients listed in Table 2 of \citet{planck2017-LIV}. For {\tt S-PASS} data this correction is negligible due to the narrowness of the bandpass.  We also subtract from the $C_\ell$ the polarization amplitude of the CMB in the corresponding multipole bin, for both $E$ and $B$-modes, taking as reference CMB values the ones coming from the {\tt Planck} 2015 $\Lambda$CDM best fit with $r=0$ \citep{planck2015-XIII}. As we stressed already, since in the simulation setup described in Section \ref{Section:sims} we include the CMB signal, our error bars on spectra include the contribution of cosmic variance. \par

\subsubsection{Results on data}
\label{cross_fit_data}
With the setup described in the previous Section we compute the fit of the model in Equation \ref{SED1} to the data. The fit is performed using a Marcovian Chain Monte Carlo (MCMC) algorithm, implemented with the python programming language. \par

Results of the fit are shown in Figures \ref{fit_nocross}, \ref{all_SED} and Table \ref{fit_results}. In the majority of the considered cases, data are well fitted by the simple power law model adopted, with $4$ out of $72$ cases having a PTE value of the fit below 5\%. The points with PTE$<0.05$ are for multipoles $\ell>100$, where the signal to noise ratio for {\tt WMAP} and {\tt Planck} data is low and where {\tt S-PASS} maps may include contamination by point sources. \par

We are able to constrain the spectral index $\beta_s$ for all the fitted SEDs, recovering an average value $\beta_s=-3.22\pm0.08$, corresponding to weighted mean and dispersion of the points. The value of $\beta_s$ appears to be rather constant across the different sky regions, and as a function of the angular scale. Moreover, we do not find any significant difference in the synchrotron behavior for $E$ and $B$ modes.\par

We stress the remarkable results we are achieving here. First, despite of fitting in an unprecedentedly wide range of frequencies, adding the {\tt S-PASS} channel at 2.3 GHz to the typical interval of frequencies of CMB experiments, the recovered value of $\beta_s$ is in agreement with what already found in previous studies of synchrotron SED from satellites at frequencies more than a factor 10 higher \citep{Fuskeland14, planck2017-LIV}. Second, constancy of the SED in angular bins, measurable for the first time, thanks to the high signal to noise ratio of the {\tt S-PASS} data, seems a characteristic feature of synchrotron, stable for all sky fractions we consider. 

\subsubsection{Validation on simulations}
\label{cross_fit_sims}
We check the consistency of our fitting procedure on simulations. In particular, we use the set of one hundred realizations of signal+noise maps described in Section \ref{Section:sims}.\par

We fit the model of Equation \ref{SED1} to the mean value of the spectra obtained from the simulations. We stress that the synchrotron signal in our simulations is rigidly rescaled at the different frequencies considering a constant spectral index $\beta_s=-3.1$, and any possible cause of de-correlation among frequencies (other than noise) is excluded. Therefore, differently to what we have described previously for data, we fit the SED model on simulations, considering the full set of ten frequencies, including therefore also the cross spectra among  {\tt S-PASS} and {\tt WMAP}/{\tt Planck}.\par

Results of the fits are shown in Figure \ref{fit_sims}. We are able to recover the input value of the $\beta_s$ parameter in all the considered cases. \par
 
\subsection{Correlation between {\tt S-PASS} and {\tt WMAP}/{\tt Planck} polarization maps}
\label{Sec:decorr}
In the upper panel of Figure \ref{all_SED} we show the amplitude of $EE$ and $BB$ spectra we get from data ($D_{\ell}$), for all the multipole bins and sky masks, together with the best fit curves. The lower panel of the same figure shows the residuals of each fit. In all these plots the points inside the grey shaded area come from the cross correlation between the {\tt S-PASS} polarization maps and the other three maps at higher frequencies from {\tt WMAP} and {\tt Planck} data.\par

As described in the previous Section, we did not consider these points while performing the fit. The reason of excluding them appears clear while looking at the residuals: the majority of them show a lack of power with respect to the best synchrotron SED model. In particular, residuals can be more than 4$\sigma$ away from the best fit curve for the largest masks. On the other hand, at high latitudes deviations are generally  within $2\sigma$, and therefore not statistically significant. This indicates that {\tt S-PASS} and {\tt WMAP/Planck} maps do not properly correlate in the sky regions close to the Galactic plane and that, here, some kind of mechanism causing de-correlation is present.\par

In general, de-correlation may originate either from instrumental effects or physical motivations. In our case, systematics effects can not be the primary source since they would cause stronger de-correlation where the signal is weaker, i.e. at high Galactic latitudes, which is opposite to what we find. For WMAP and Planck, residual intensity-to-polarization leakage and $1/f$ noise contribution can be present. However, the low signal-to-noise ratio makes it difficult to assess the proper level of those signals and estimate their effects on de-correlation. S-PASS has low systematics, unlikely to generate such effects. In particular, the leading spurious signal comes from residual ground emission, but it is too small (less than 3\% of the signal in the low emission areas at high latitudes, \citet{carretti_etal_in_preparation}) to cause significant effects on spectra and, in turn, on de-correlation. \par

A second, more likely, cause for de-correlation is Faraday rotation at 2.3 GHz which is present, as we already noticed, in those sky regions close to the Galactic plane, while its effect is negligible at {\tt WMAP} and {\tt Planck} frequencies. The fact that residuals, for the three cross-correlation points, are larger on those sky masks including also lower Galactic latitudes, represents a hint that Faraday rotation plays an important role in causing de-correlation, at least on these large sky regions. Nevertheless, the lack of independent data estimating the level of Faraday rotation does not allow a thorough characterization of the effect.\par

Lastly, it is important to notice that we cross-correlate maps at frequencies which are far away from each other, and we are fitting the data with a simple power-law model, having a single spectral index $\beta_s$. Any deviation from this approximated model, caused for example by a frequency or a spatial variation of the spectral index can cause large residuals, especially visible in the cross-correlation of data at distant frequencies. The possibility that de-correlation is actually due to the variation of the underlying physical properties of synchrotron would have important implications for $B$-mode experiments targeting cosmological GWs. It would imply that these experiments need to design focal planes monitoring synchrotron at frequencies close to the one of minimum foreground contamination, for properly mapping the foreground contamination and subtract it out with low residuals on the CMB signal.\par

A quantitative assessment of de-correlation and its physical properties, though, will have to wait for other independent data at intermediate frequencies. Several observations are ongoing in the northern \citep[{\tt QUIJOTE, C-BASS, LSPE}]{2017MNRAS.464.4107G,2018arXiv180504490J, 2012SPIE.8446E..7AA} and southern \citep[{\tt C-BASS, CLASS}]{2016AAS...22830104P, CLASS} hemispheres which will allow to further characterize this effect.

\begin{figure*}[!t]
\centering
\includegraphics[width=16.5 cm]{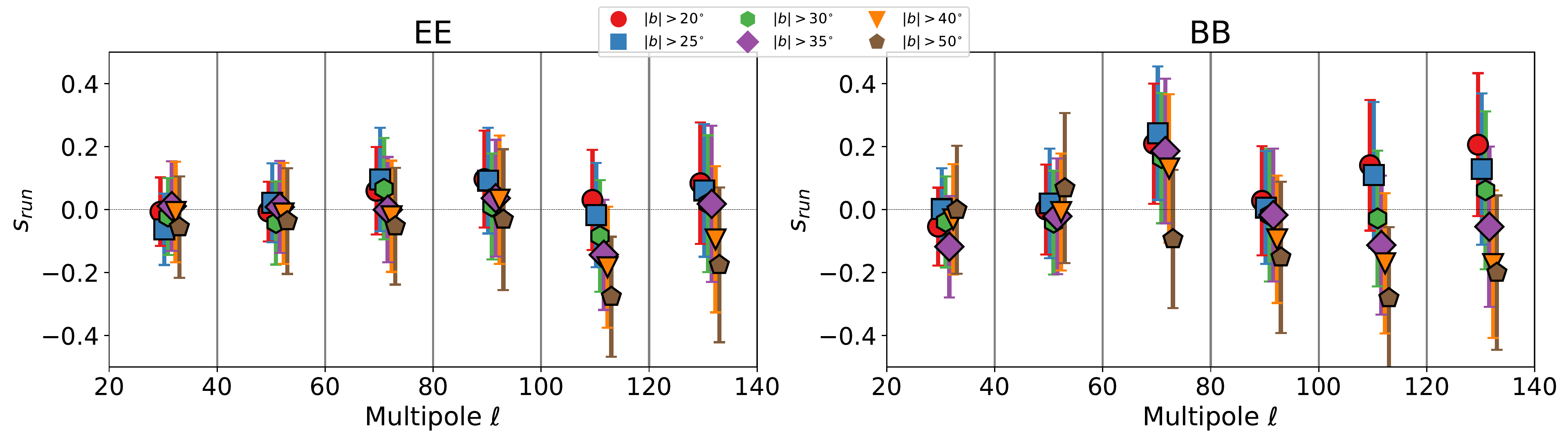}
\caption{Constraint on the $s_{run}$ parameter describing the curvature of the synchrotron spectral index in polarization.}
\label{fit_srun}
\end{figure*}

\begin{figure}[!h]
\centering
\includegraphics[width=9 cm]{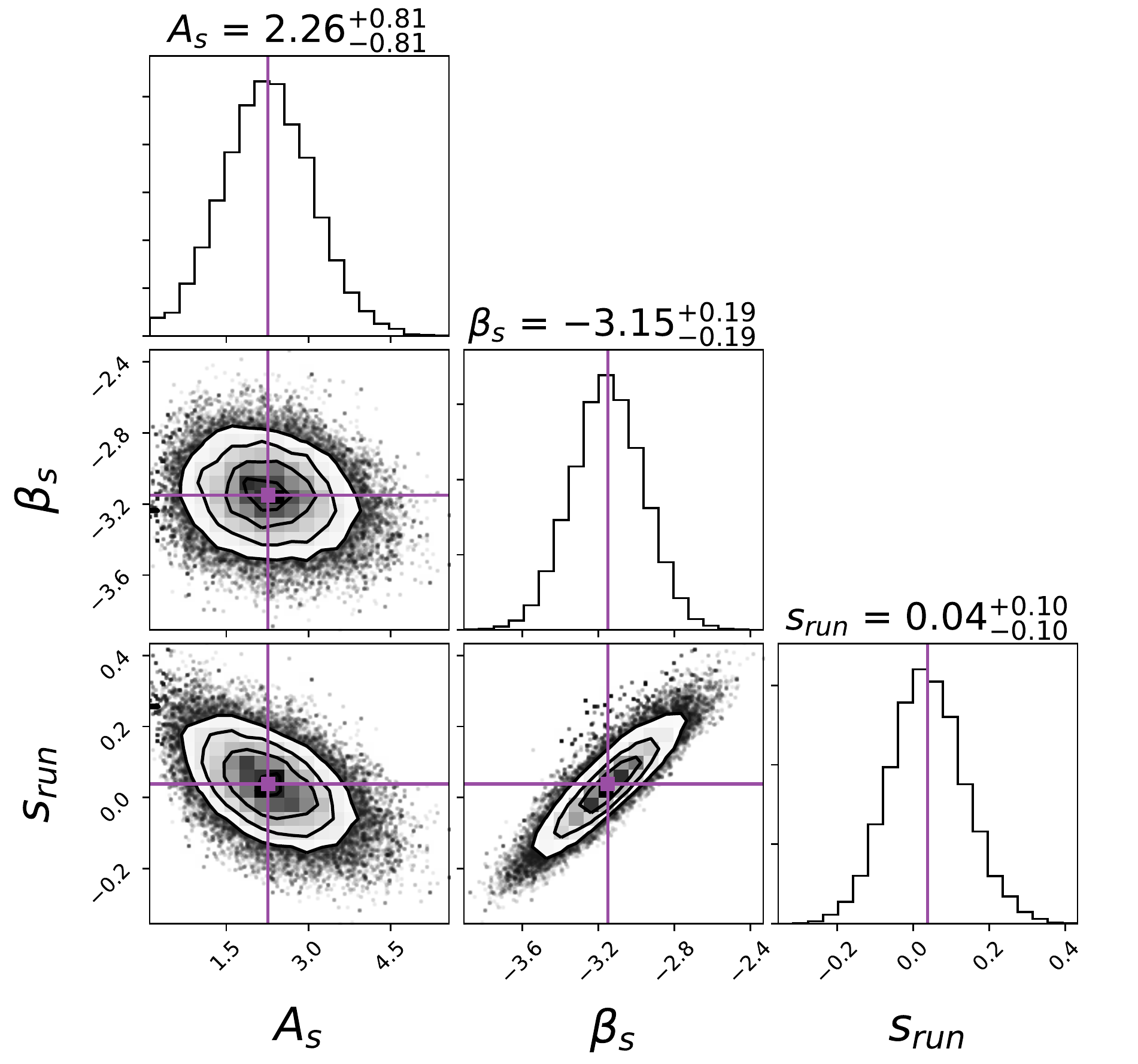}
\caption{Posterior distribution of the parameters obtained  by fitting the model in \ref{SED2} to $BB$ power spectra of {\tt S-PASS}, {\tt Planck} and {\tt WMAP}  in the multiple bin at $\ell\simeq50$ and for the iso-latitude mask with $|b|>35^{\circ}$. The fit has been computed by applying a Gaussian prior on the synchrotron spectral index with $\beta_s=-3.13\pm0.13$.}
\label{corner_srun}
\end{figure}

\subsection{Constraints on synchrotron curvature}
As aforementioned, by fitting the data with a simple power law model, in the frequency range 2.3-33 GHz, we retrieve an average value for the spectral index of synchrotron in agreement with constraints obtained considering only higher frequency data. This result, therefore, seems to exclude the possibility of having a strong curvature of the synchrotron SED with frequency, in polarization. In order to  better quantify this statement, we fit the same set of data described in Section \ref{Sec:4.1} with a model that also includes a curvature of the synchrotron spectral index. In particular we fit the $C_{\ell}(\nu_1 \times \nu_2)$ amplitude of power spectra as follows:
\begin{equation}
C_{\ell}(\nu_1 \times \nu_2) = A_s \left(\frac{\nu_1}{\nu_0}\right)^{\beta_s+s_{run}\log(\nu_1/\nu_{0})}\left(\frac{\nu_2}{\nu_0}\right)^{\beta_s+s_{run}\log(\nu_2/\nu_{0})}. 
\label{SED2}
\end{equation}
allowing, therefore, a logarithmic steepening or flattening of the synchrotron spectrum with frequency, through the $s_{run}$ parameter \citep{2007ApJ...665..355K, 2012ApJ...753..110K}.\par

In performing the fit we apply a Gaussian prior on the $\beta_s$ parameter with $\beta_s=-3.13\pm0.13$. This prior is the same adopted in \citet{planck2017-LIV} and has been obtained as the mean value of the $\beta_s$ computed from {\tt Planck} and {\tt WAMP} data in sky regions with $f_{sky}$ ranging from 24\% to 71\% and multipole bins from $\ell=4$ to $\ell=160$. The reason for applying this prior is that the $\beta_s$ and $s_{run}$ parameters are highly correlated. Moreover, the constraint on $\beta_s$ from {\tt WMAP} and {\tt Planck} data, on our sky masks, is too weak to allow the fit of both parameters simultaneously, without any prior. Since the prior on the $\beta_s$ comes from {\tt WMAP} and {\tt Planck} data, in this case we fix the pivot frequency $\nu_0$  at 23 GHz.\par

Figure \ref{fit_srun} and Table \ref{fit_srun_tab} report the resulting constraints on the $s_{run}$ parameter, in each multipole bin and sky maps. In all the cases the $s_{run}$ parameter is compatible with zero at $1.5\sigma$ at most, with error bars ranging from 0.07 to 0.14. In Figure \ref{corner_srun} we show the posterior distribution from the MCMC fit, in the particular case at $\ell\simeq50$, iso-latitude  mask with $|b|>35^{\circ}$ and $BB$ spectrum.

\begin{table*}[!t]
      \caption{Mean value and $1\sigma$ error of the $s_{run}$ parameter in the different multiple bins and considered sky mask for $E$ and $B$-modes.}
         \label{fit_srun_tab}
     $$ 
       \begin{threeparttable}
         \begin{tabular}{lccccccc} 
            \hline\hline
            \noalign{\smallskip}\noalign{\smallskip}
          && $|b|>20^{\circ}$ &   $|b|>25^{\circ}$&  $|b|>30^{\circ}$ &  $|b|>35^{\circ}$ & $|b|>40^{\circ}$ & $|b|>50^{\circ}$ \\           
            \noalign{\smallskip}
             \hline
            \noalign{\smallskip}
              $f_{sky} $ &           &0.30 & 0.26 & 0.22 & 0.19 & 0.16& 0.10       \\
             \noalign{\smallskip}                                   
            \hline
            \noalign{\smallskip}
& $20\leq\ell\leq39$ & 0.02$\pm$0.08 & -0.01$\pm$0.08 & 0.02$\pm$0.08 & 0.05$\pm$0.08 & 0.04$\pm$0.09 & -0.0$\pm$0.09 \cr
& $40\leq\ell\leq59$ & 0.01$\pm$0.07 & 0.03$\pm$0.08 & 0.0$\pm$0.08 & 0.03$\pm$0.08 & 0.03$\pm$0.09 & 0.01$\pm$0.09 \cr
$s_{run}^{EE}$ & $60\leq\ell\leq79$ & 0.05$\pm$0.08 & 0.07$\pm$0.09 & 0.04$\pm$0.09 & 0.03$\pm$0.09 & 0.03$\pm$0.09 &  0.0$\pm$0.10 \cr
& $80\leq\ell\leq99$ & 0.07$\pm$0.09 & 0.07$\pm$0.09 & 0.02$\pm$0.09 & 0.04$\pm$0.10 & 0.05$\pm$0.11 & 0.02$\pm$0.12 \cr
& $100\leq\ell\leq119$ & 0.02$\pm$0.09 & 0.01$\pm$0.09 & -0.01$\pm$0.09 & -0.05$\pm$0.09 & -0.07$\pm$0.10 & -0.11$\pm$0.10 \cr
& $120\leq\ell\leq139$ & 0.08$\pm$0.10 & 0.07$\pm$0.11 & 0.06$\pm$0.12 & 0.05$\pm$0.12 & -0.01$\pm$0.12 & -0.06$\pm$0.13 \cr
 \noalign{\smallskip}
\hline\hline
\noalign{\smallskip}
& $20\leq\ell\leq39$ & 0.01$\pm$0.08 & 0.03$\pm$0.08 & 0.02$\pm$0.08 & -0.0$\pm$0.09 & 0.02$\pm$0.09 & 0.04$\pm$0.10 \cr
& $40\leq\ell\leq59$ & 0.04$\pm$0.10 & 0.07$\pm$0.09 & 0.03$\pm$0.09 & 0.04$\pm$0.10 & 0.04$\pm$0.1 & 0.07$\pm$0.12 \cr
$s_{run}^{BB}$ & $60\leq\ell\leq79$ & 0.15$\pm$0.10 & 0.17$\pm$0.11 & 0.13$\pm$0.11 & 0.14$\pm$0.12 & 0.11$\pm$0.12 & -0.0$\pm$0.12 \cr
& $80\leq\ell\leq99$ & 0.05$\pm$0.09 & 0.05$\pm$0.10 & 0.03$\pm$0.10 & 0.03$\pm$0.11 & -0.02$\pm$0.11 & -0.06$\pm$0.12 \cr
& $100\leq\ell\leq119$ & 0.12$\pm$0.10 & 0.11$\pm$0.11 & 0.04$\pm$0.11 & -0.01$\pm$0.11 & -0.03$\pm$0.12 & -0.1$\pm$0.12 \cr
& $120\leq\ell\leq139$ & 0.15$\pm$0.12 & 0.11$\pm$0.13 & 0.08$\pm$0.14 & 0.01$\pm$0.13 & -0.04$\pm$0.13 & -0.06$\pm$0.14 \cr

\noalign{\vskip 3pt\hrule\vskip 5pt}
         \end{tabular}
     \end{threeparttable}
     $$ 
      \end{table*}


\begin{figure}[!h]
\includegraphics[width=8.5 cm]{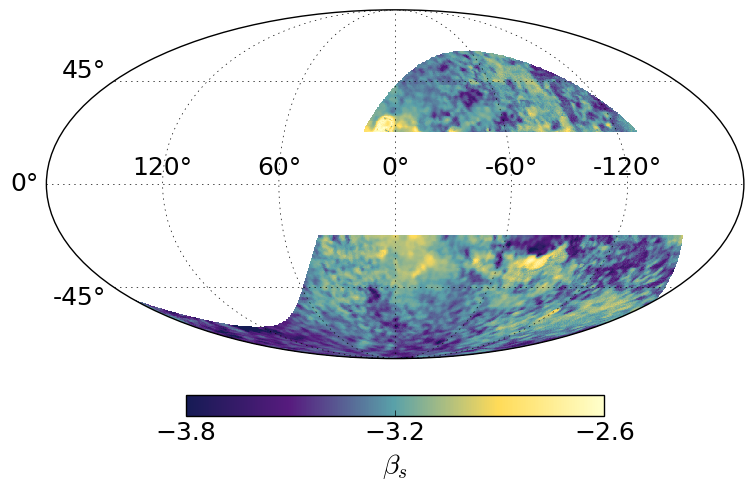}\\
\includegraphics[width=8.5 cm]{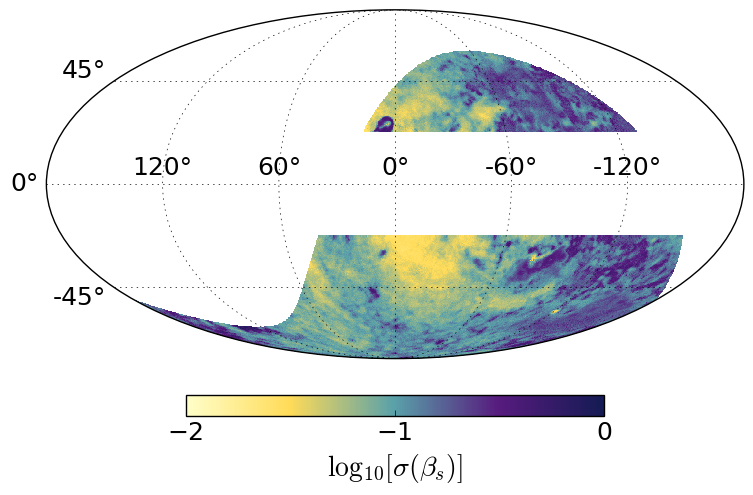}\\
\includegraphics[width=8.5 cm]{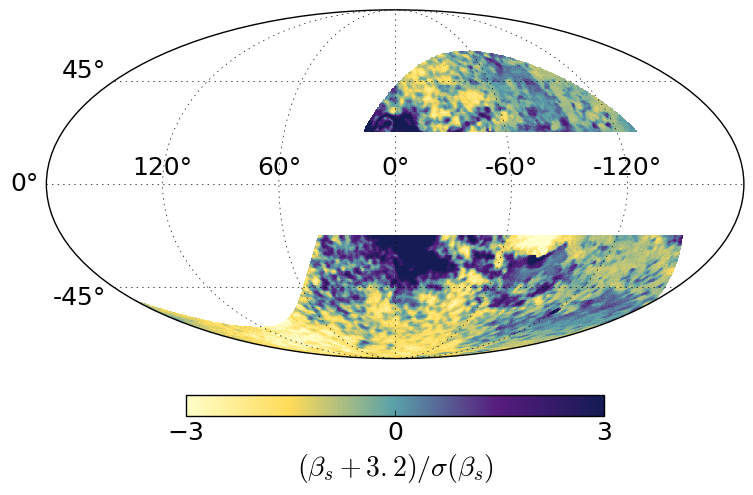}\\
\caption{\label{maps_beta} Upper panel: synchrotron spectral index map derived as described in the text. 
Middle panel: $1\sigma$ uncertainty on $\beta_s$. Lower panel: significance of the spectral index variation with respect to $\beta_s=-3.2$, corresponding to the value at which the distribution of the $\beta_s$ on map peaks (see Figure \ref{beta_hist}). 
Note that colors are saturated for visualization purposes. The complete range of values is: $-4.4\leq\beta_s\leq-2.5$, $-1.6\leq\log_{10}[\sigma(\beta_s)]\leq0.03$, $-6\leq S/N\leq20$}
\end{figure}

\begin{figure}[!h]
\centering     
\includegraphics[width=7 cm]{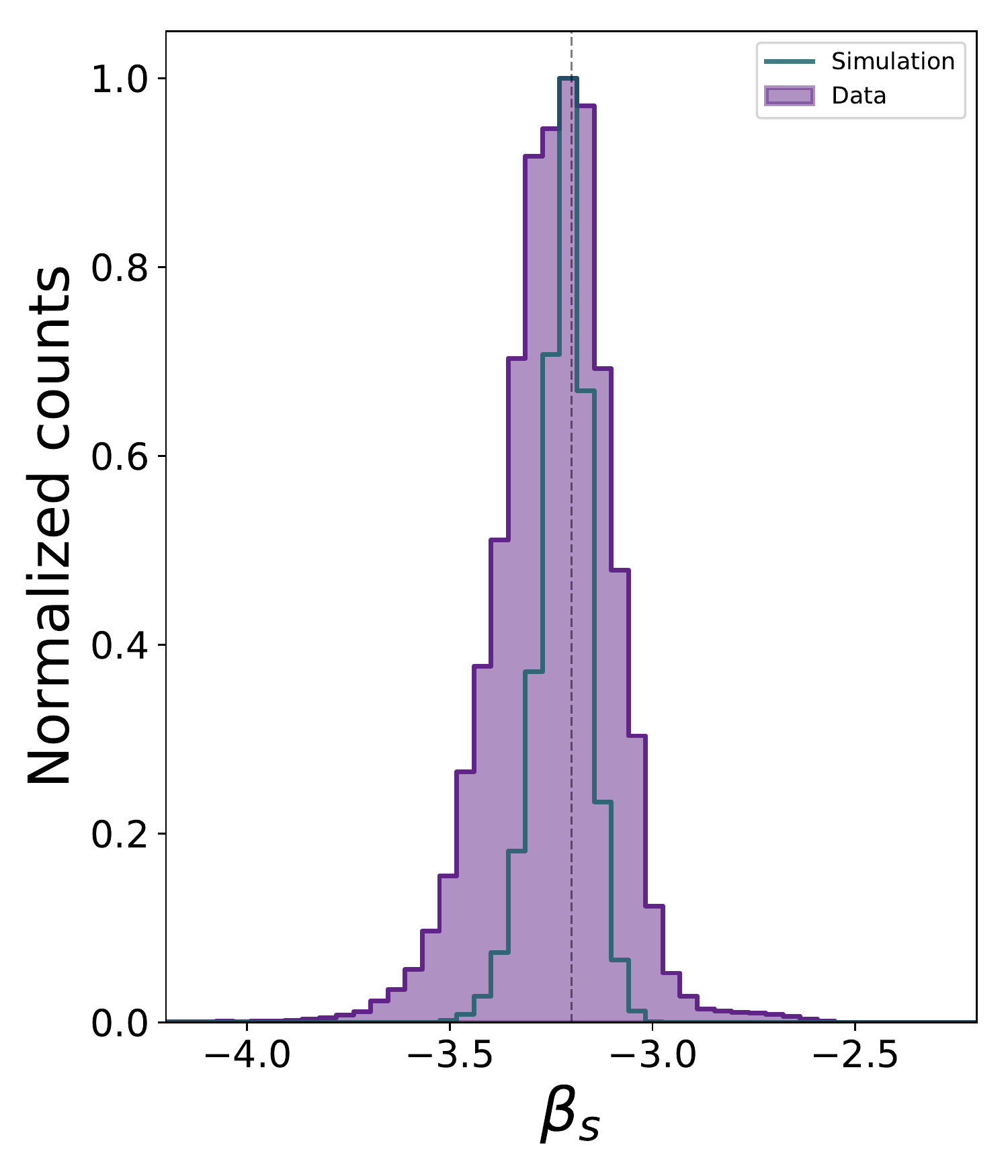}
\caption{Comparison of the normalized histograms of the synchrotron spectral index map obtained from data (indigo) and simulation (cyan line). The dashed line is at $\beta_s=-3.2$, where the $\beta_s$ distribution peaks and also represents the reference value of the simulated case. }
\label{beta_hist}
\end{figure}

\begin{figure}[!h]
\centering     
\includegraphics[width=9 cm]{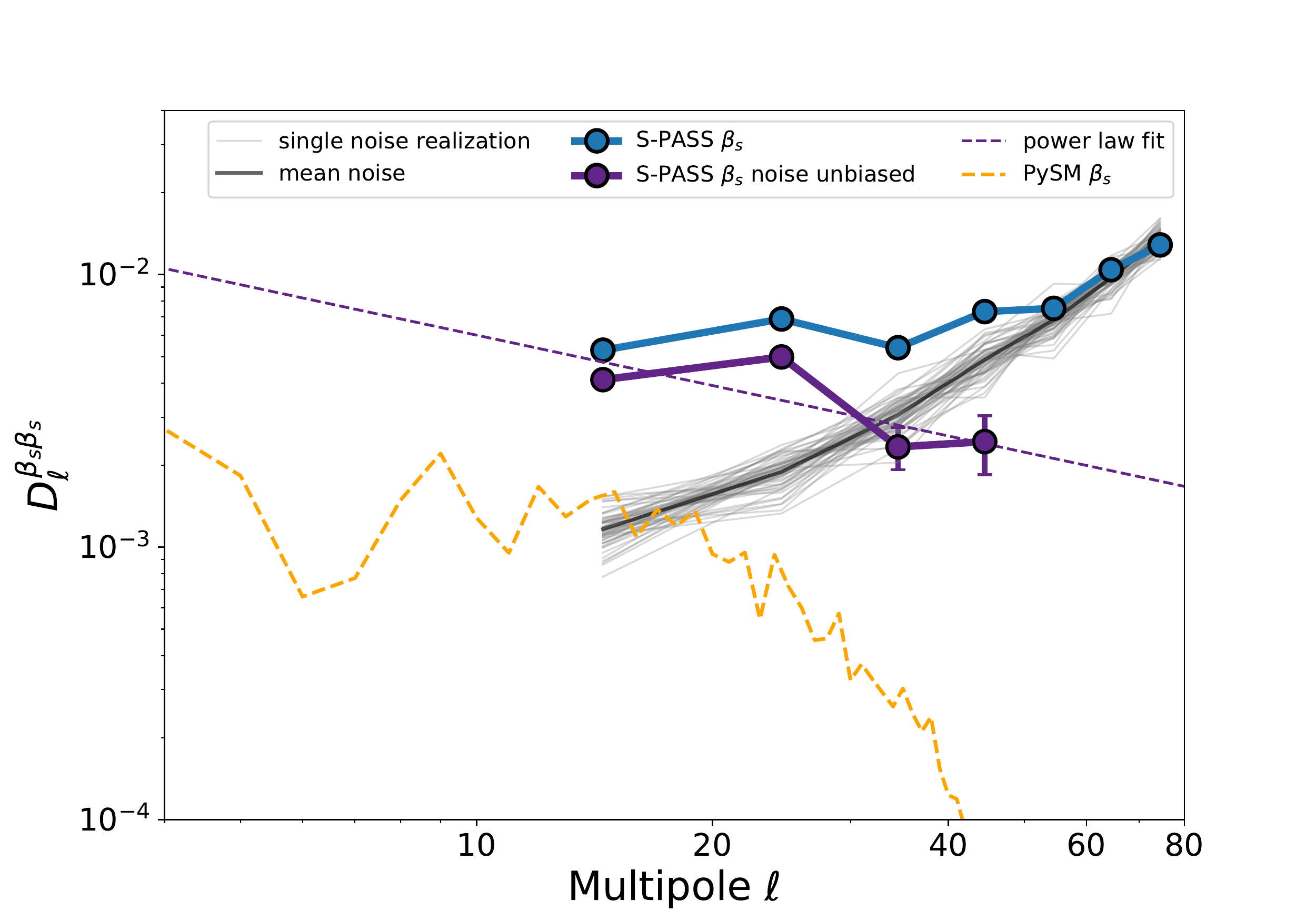}
\caption{Angular power spectrum of the $S-PASS$ $\beta_s$ map before (blue) and after (purple) correction for the contribution of noise. For a complete description of the Figure see text in Section \ref{Sec:beta_spectrum}.}
\label{beta_spectrum}
\end{figure}

\section{Polarized synchrotron spectral index map}
\label{Section:5}

In the previous section we describe how we used {\tt S-PASS}, {\tt WMAP} and {\tt Planck} data to estimate the synchrotron spectral index in the harmonic domain. In this Section we estimate $\beta_s$ by performing a pixel based analysis for deriving the corresponding map. \par

\subsection{Procedure}
\label{Subsection:procedure}
We evaluate the synchrotron spectral index considering again the simple power law model and fitting the data in the pixel domain. As before, we consider the {\tt S-PASS}, {\tt WMAP} $K$, $Ka$ and {\tt Planck-LFI} 30 GHz polarization maps. Contrary to what we have done with power spectra, we fix the amplitude of synchrotron at the value measured at 2.3 GHz in the {\tt S-PASS} data and fit only for the spectral index $\beta_s$. We perform this analysis on the sky region observed by   {\tt S-PASS} at $|b|>20^{\circ}$.\par  

In the pre-processing, we smooth all the maps at the same angular resolution of $2^{\circ}$. We fit for $\beta_s$ in each pixel, after degrading the input maps at $N_{side}=256$. \par  

We perform the fit on the polarized intensity maps $P$ (with $P=\sqrt{Q^2+U^2}$), in order to avoid effects coming from possible rotation in the polarization angle with frequency. Nevertheless, total polarization maps are positively biased, due to the presence of noise. This bias is negligible for {\tt S-PASS} which, at the angular resolution of $2^{\circ}$, has $S/N>5$ everywhere on both stokes $Q$ and $U$ maps. On the contrary, the bias is important for {\tt WMAP} and {\tt Planck}, which have lower signal to noise. If not properly taken into account, the noise bias can therefore cause a shift of the recovered spectral index $\beta_s$ towards higher values (flatter spectra).\par 

In order to obtain an unbiased estimate, we include the presence of noise while preforming the fit. In particular, in each pixel we minimize the following function:
\begin{equation}
f(\beta_s) = \sum_{\nu_i} (\tilde{P}_{\nu_i}-P_{\nu_i})^2,
\label{chi2_beta_map}
\end{equation}
where $\nu_i\in[23, 28.4, 33]$ GHz and $P_{\nu_i}$ is the total polarization amplitude observed on maps at  frequency $\nu_i$. $\tilde{P}_{\nu_i}$ is computed starting from the {\tt S-PASS} polarization maps as:
\begin{equation}
\tilde{P}_{\nu_i} = \sqrt{\left[Q_{2.3}\left(\frac{2.3}{\nu_i}\right)^{\beta_s}+n^Q_{\nu_i}\right]^2+\left[U_{2.3}\left(\frac{2.3}{\nu_i}\right)^{\beta_s}+n^U_{\nu_i}\right]^2},
\label{beta_map_model}
\end{equation}
where $Q_{2.3}$ and $U_{2.3}$ represent the amplitude of {\tt S-PASS} Stokes $Q$ and $U$ maps respectively, while $n^Q_{\nu_i}$ and $n^U_{\nu_i}$ are a random realization of white noise, for $Q$ and $U$, at the frequency $\nu_i$. These random realizations are computed from the variance map of the noise at $2^{\circ}$, for each different frequency $\nu_i$.\footnote{The variance maps are obtained from a hundred noise realizations, generated as described in Section \ref{Section:sims}, after having smoothed them to get the angular resolution of $2^{\circ}$ ($FWHM$).} $\tilde{P}_{\nu_i}$ represents, therefore, our model of the total polarization amplitude at the frequency $\nu_i$, and includes also the noise bias. We stress again that, given our model, the only free parameter in performing the minimization of the function in Equation \ref{chi2_beta_map} is the spectral index $\beta_s$ while the synchrotron amplitude is fixed at the value we observe in {\tt S-PASS} maps (that we consider to be noiseless). Obviously, in order to take into account properly the effect of noise, we have to perform the fit multiple times, randomizing on the noise realizations $n^Q_{\nu_i}$ and $n^U_{\nu_i}$. In particular, for each pixel, we compute the fit a hundred times getting the final value and the statistical error on $\beta_s$ as mean and standard deviation of these estimated values. 

\subsection{$\beta_s$ map and errors}
\label{subsection:beta_maps}
Figure \ref{maps_beta} shows the recovered spectral index map and corresponding errors (upper and middle panel, respectively). In the error budget we include both the statistical error (obtained as described in the previous section) and the error coming from the uncertainty in the {\tt S-PASS} calibration, which is  5\% on map. The two kinds of uncertainty are added together in quadrature to obtain the final value of $\sigma(\beta_s)$ in each pixel. In the maps, values of the spectral index vary in the range $-4.4\leq\beta_s\leq-2.5$, with $0.02\leq\sigma(\beta_s)\leq1.06$. In about 46\% of the analyzed sky the spectral index value is recovered with a total uncertainty below 0.1. \par

Figure \ref{beta_hist} shows the histogram of the recovered $\beta_s$ on map. This distribution peaks around $\beta_s\simeq-3.2$, with an average value $\beta_s=-3.25\pm0.15$. These numbers are in agreement with the value we find fitting for the synchrotron SED in harmonic space (see Section \ref{cross_fit_data}), and are also consistent with the result obtained by \citet{2010MNRAS.405.1670C} who, using polarization maps at 1.4 and 23 GHz found a spectral slope at high Galactic latitudes that peaked around -3.2. We compare the distribution of the $\beta_s$ with the one obtained from simulations. In particular, in the simulated case, we extrapolate the {\tt S-PASS} polarization maps to {\tt WMAP} and {\tt Planck} frequencies, considering a fixed spectral index $\beta_s^{SIM} = -3.2$. After properly adding noise on maps at the different frequencies, and smoothing them at the angular resolution of $2^{\circ}$, we perform the fit with the procedure described in Section \ref{Subsection:procedure}. The histogram of the recovered values of the spectral index on map is also shown in Figure \ref{beta_hist}. We first observe that the histogram peaks at $\beta_s\simeq-3.2$, showing that our fitting procedure recovers a spectral index map which is unbiased. Moreover, we notice that the $\beta_s$ distribution for the simulated case appears to be narrower when compared to the one from data. This indicates that the spread of $\beta_s$ we find is a real feature of the sky emission and not only due to noise.\par

In the lower panel of Figure \ref{maps_beta} we also show an estimate of the significance of the recovered variation of the spectral index with respect to the reference value $\beta_s=-3.2$ in each pixel, computed as $S = (\beta_s+3.2)/\sigma(\beta_s)$. We find that $|S|>2$ in about 12\% of the analyzed sky.\par

We stress here that the synchrotron spectral index map obtained in this work is the first available one computed using only polarized data, and represents, therefore, an important tool to generate realistic simulations of the polarized synchrotron sky\footnote{The spectral index map is available upon request to be sent to Nicoletta Krachmalnicoff (\href{mailto:nkrach@sissa.it}{nkrach@sissa.it})}.

\subsection{Angular power spectrum of the $\beta_s$ map}
\label{Sec:beta_spectrum}
We compute the angular power spectrum of the $\beta_s$ map, in order to get information about the amplitude of  the spectral index variations as a function of the angular scale. \par

We compute the $\langle\beta_s\beta_s\rangle$ auto-power spectrum on the full sky region covered by our map (about 30\% of the sky), in the multipole range 10-80 with $\Delta\ell=10$ and correcting for a Gaussian beam window function with $FWHM=2^{\circ}$. The spectrum is shown in Figure \ref{beta_spectrum} (cyan curve). We correct this spectrum for the noise bias, i.e. the the contribution coming from $\beta_s$ fluctuations due the presence of noise on the {\tt WMAP} and {\tt Planck} maps. Estimating this contribution is not trivial, since the level of fluctuations on the $\beta_s$ map due to noise depends on the value of the $\beta_s$ itself. We therefore rely on the following procedure:
\begin{itemize}
\item we start from the {\tt S-PASS} total polarization map that we extrapolate at {\tt WMAP/Planck} frequencies using our $\beta_s$ map;
\item we add a noise realization representative of {\tt WMAP/Planck} noise to the extrapolated maps (following the procedure described in Section \ref{Section:sims});
\item on this set of simulated maps we estimate the value of the spectral index $\beta_s^*$ with the procedure used for data;
\item we compute the power spectrum of the $\beta_{diff} = (\beta_s^*-\beta_s)$ map (thin grey lines on Figure \ref{beta_spectrum});
\item we repeat this procedure a hundred times changing the noise realizations;
\item we evaluate the noise bias as the mean of the obtained one hundred spectra (black line on Figure \ref{beta_spectrum});
\item the unbiased $\beta_s$ spectrum is obtained by subtracting this mean curve to the spectrum of $\beta_s$;
\item error bars on the unbiased spectrum are obtained as the standard deviation of the hundred noise spectra.
\end{itemize} \par

The unbiased spectrum is shown in Figure \ref{beta_spectrum} in purple, for the four multiple bins not compatible with zero. In order to extrapolate the amplitude of fluctuations at all angular scales we fit these points with a power law model with $C_{\ell}\propto\ell^{\gamma}$, finding a value of $\gamma=-2.6\pm0.2$ (dashed purple line). We also compare our results with the power spectrum of the synchrotron spectral index map (computed on the same 30\% sky region of our analysis) currently used in the sky modeling for many CMB experiments, i.e. the map included in the PySM simulation package \citep{2017MNRAS.469.2821T}), shown in orange on Figure \ref{beta_spectrum}. We stress that this map was obtained combining the first {\tt WMAP} polarization data with the Haslam total intensity ones at 408 MHz \citep{1981A&A...100..209H}, considering a model for the Galactic magnetic field, and it includes variations of the synchrotron spectral index only on scale larger then $\sim7^{\circ}$ \citep{2008A&A...490.1093M}. The extrapolation of our results to low multipoles leads to fluctuations of the spectral index with amplitude about 3 times larger (at the spectrum level) than those in the PySM $\beta_s$ map.


\begin{figure*}[!t]
\centering
\includegraphics[width=18 cm]{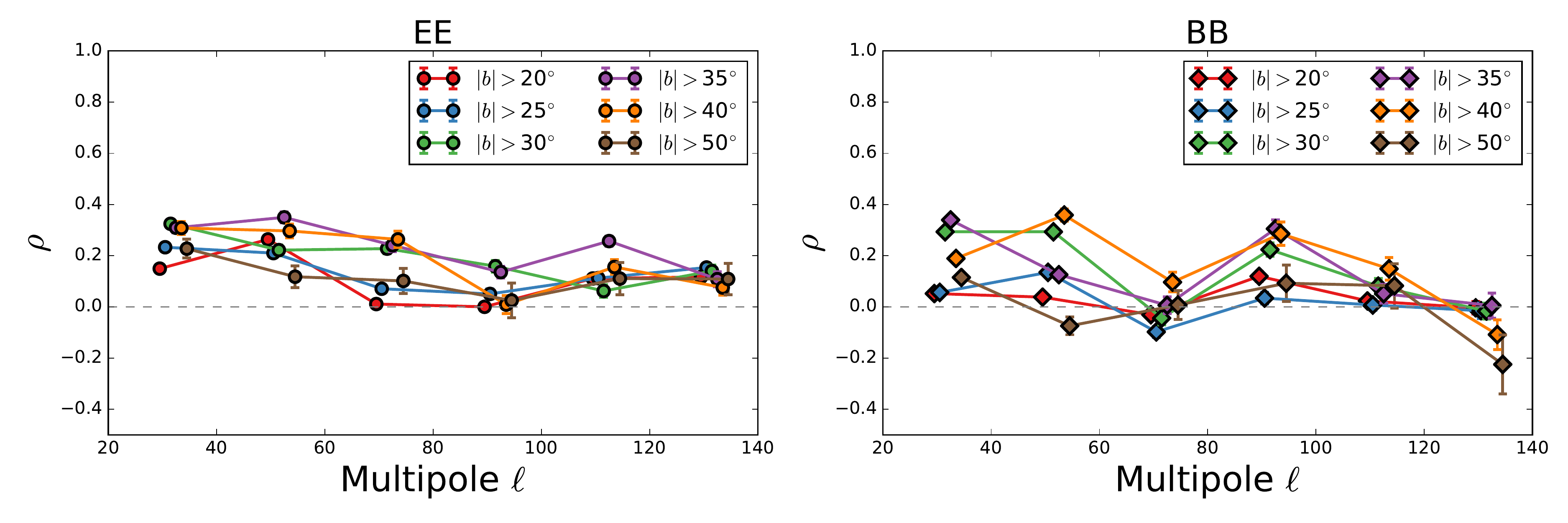}
\caption{Correlation coefficient for thermal dust and synchrotron computed as in Equation (\ref{rho_eq}), using {\tt S-PASS} and {\tt Planck} 353 GHz polarization maps.}
\label{rho_spectrum}
\end{figure*}

\section{Spatial correlation with thermal dust emission}
\label{Section:6}

Polarized Galactic synchrotron and thermal dust signals are expected to have some degree of correlation, due to the same underlying Galactic magnetic field.\par 

This correlation has already been measured on the larger angular scales ($\ell\lesssim100$) through the computation of cross angular power spectra between {\tt WMAP} and {\tt Planck} low and high frequency channels 
(see \citet{2015JCAP...12..020C,planck2017-LIV}). In these works, the reported values for the correlation coefficient $\rho$ show a progressive decay for increasing multipoles, with $\rho\simeq0.5$ at $\ell\simeq10$ and approaching values compatible with zero at the degree angular scales. \par

In this Section we summarize the results we obtain by computing the cross correlation between the {\tt S-PASS} and {\tt Planck} maps at 353 GHz. 

\subsection{{\tt S-PASS}$\times${Planck}-353 GHz cross power spectra}
The {\tt Planck-HFI} ({\tt High Frequency Instrument}) observations of the full sky at 353 GHz represent the best measurement of the polarized thermal dust emission currently available. They have been intensively used to study 
the properties of the signal and its contamination to CMB primordial $B$-modes, and several experiments are currently relying on them to perform component separation and to isolate the CMB signal \citep{B2P} or to put upper limits on foreground emissions in the observed regions of the sky \citep{PolarBear17}\par

Therefore, in order to estimate the $\rho$ coefficient describing the correlation between synchrotron and thermal dust, we use the {\tt Planck} 353 GHz $Q$ and $U$ polarization maps, and compute the cross spectra with {\tt S-PASS} data. In particular, the value of $\rho$ is obtained as follows:
\begin{equation}
\rho_{\ell} = \frac{C_{\ell}(2.3\times353)}{\sqrt{C_{\ell}(2.3)\,C_{\ell}(353)}},
\label{rho_eq}
\end{equation} 
where the numerator contains the amplitude in a given multipole bin of the cross spectrum between {\tt S-PASS} and {\tt Planck}-353 polarization maps divided by the amplitude of the single-frequency spectra (at 2.3 and 353 GHz). In particular, as previously done, the value of $C_{\ell}(2.3)$ is obtained from the auto spectra of the {\tt S-PASS} maps (given the low level of noise at the angular scale of interest). On the contrary, we estimate 
$C_{\ell}(353)$ by computing cross spectra of independent splits obtained from the {\tt Planck} dataset at 353 GHz, in order to avoid the presence of noise bias on the spectra. The {\tt Planck} HFI maps are publicly available through the Planck Legacy Archive. As independent splits we consider the two half-mission maps. \par

We evaluate $\rho_{\ell}$ by computing the $E$ and $B$-mode spectra on the set of six iso-latitude masks described previously (Section \ref{Section:mask}), in the same multipole bins we use in the analysis of the synchrotron SED, ranging from $\ell=20$ to $\ell\simeq140$ (see Section \ref{Section:SED}), and considering maps degraded at the $N_{side}=256$ pixelization.\par

The values for $\rho$ obtained in the different masks and multipole bins for $E$ and $B$-modes are shown in Figure \ref{rho_spectrum}. The error bars are again obtained through simulations; in particular, we generate a set of one hundred signal+noise simulations at 2.3 and 353 GHz, where the signal includes synchrotron and thermal dust emissions (obtained from the {\tt Planck} FFP8 sky maps), and CMB realizations (from the best {\tt Planck} 2015 
$\Lambda$CDM model). We get the {\tt S-PASS} noise as Gaussian realizations of the variance map. For the {\tt Planck}-353 GHz channel we obtain the noise from data, following the procedure described in Section 
\ref{Section:sims}, and using the half-mission maps as splits. We compute the cross-frequency and single-frequency spectra from this set of simulations, calculate the value of $\rho$ through Equation \ref{rho_eq}, and obtain the 
error bars as the standard deviations of these simulated values. \par 

The results show a level of correlation as high as 40\% at large angular scales, and a general decaying trend at higher multipoles. It is worth noticing that the correlation is positive in particular at high latitudes; for both $E$ and $B$-modes, indicating that some de-correlating contamination, probably due to Faraday modulation, may be active at low latitudes. \par

The observed correlation level at intermediate and high latitudes is in agreement with what previously found by \citet{2015JCAP...12..020C} and \citet{planck2017-LIV}, showing at the same time that {\tt S-PASS} at intermediate and high latitudes is a good tracer of synchrotron, and that the correlation between the two emissions, especially at low multipoles, persists also when we compare very distant frequencies, such as for {\tt S-PASS} and {\tt Planck}-353 GHz channels.


\section{Contamination to CMB polarization}
\label{Section:7}

\begin{figure*}[!t]
\centering
\includegraphics[width=\textwidth]{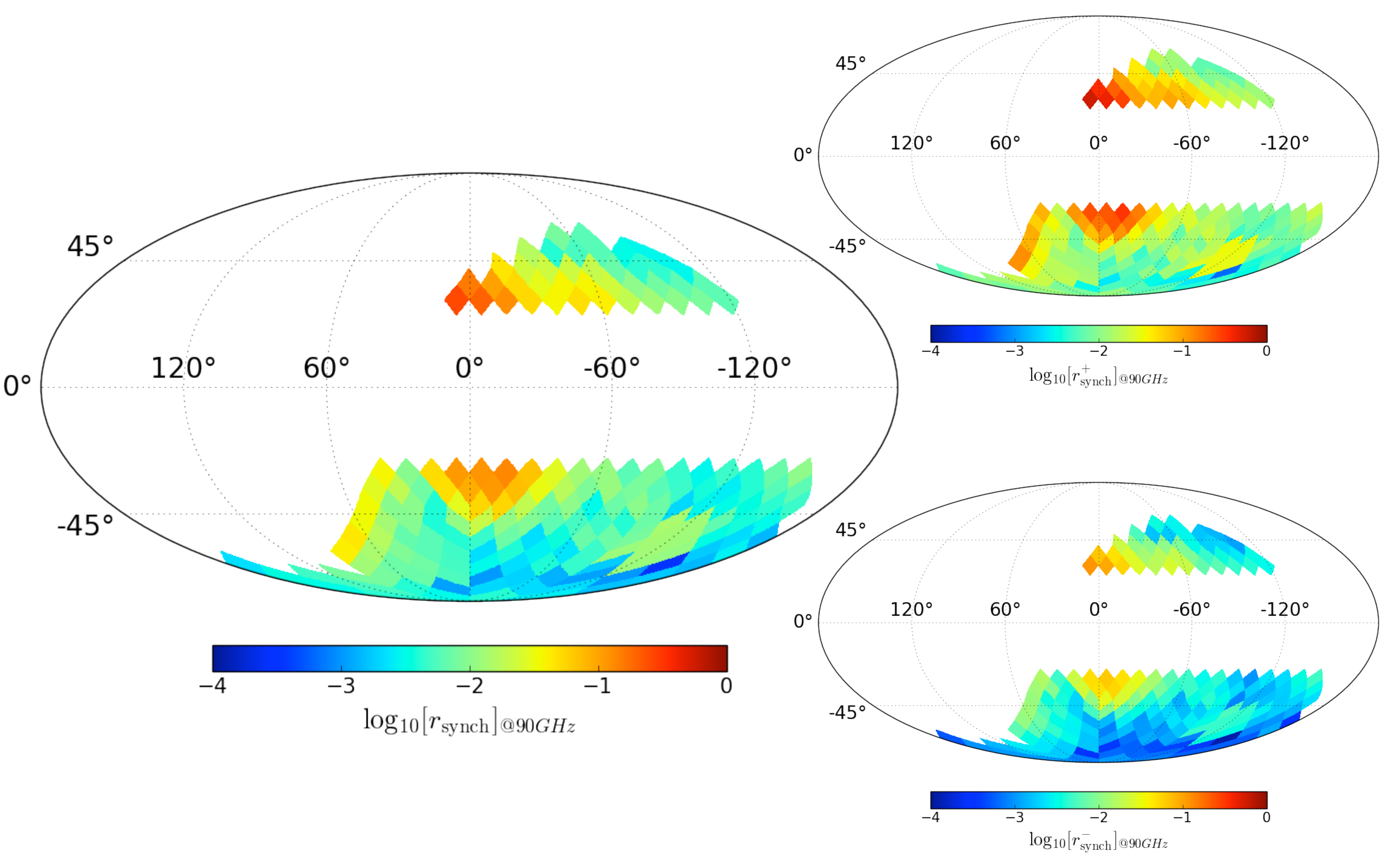}\\
\includegraphics[width=6cm]{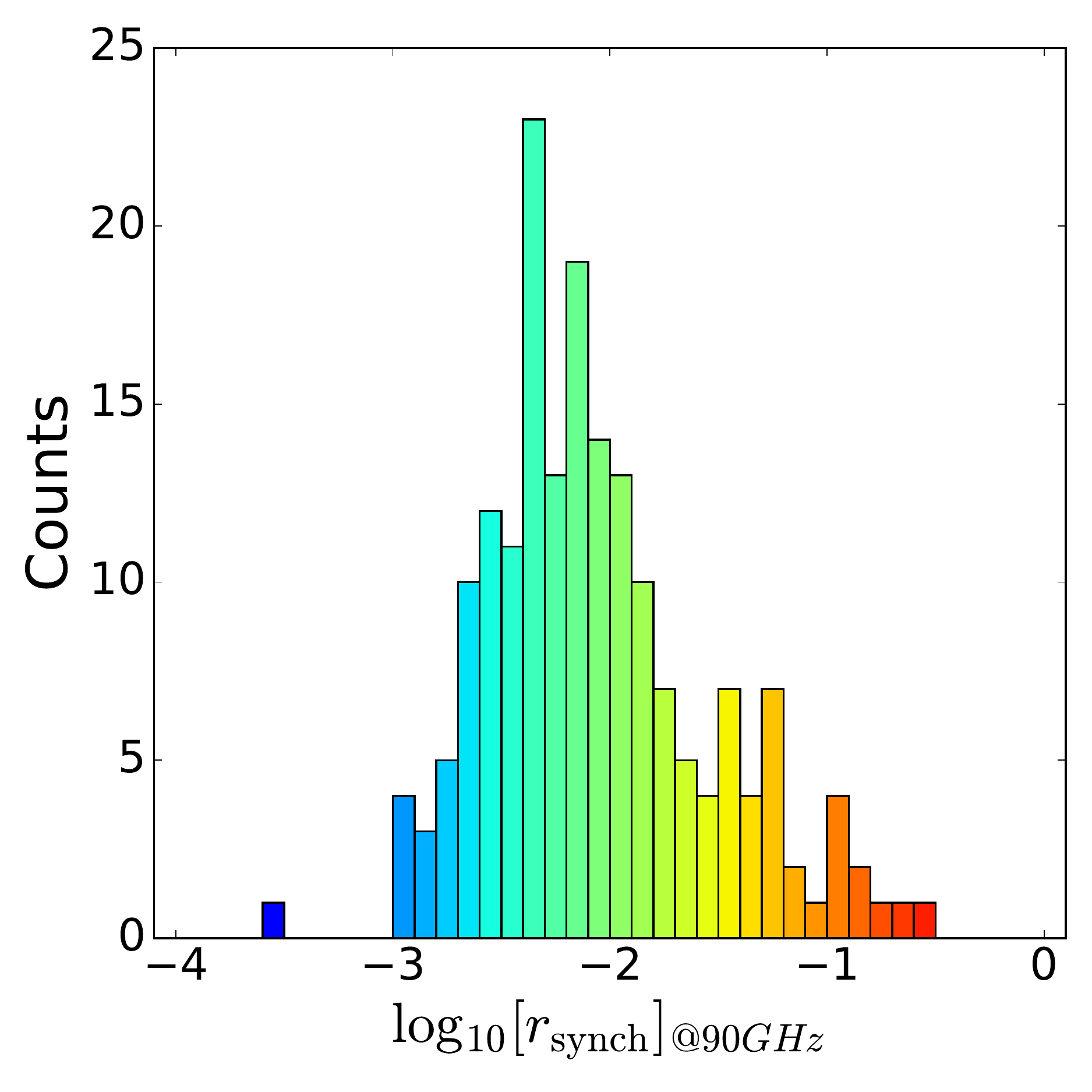}
\includegraphics[width=6cm]{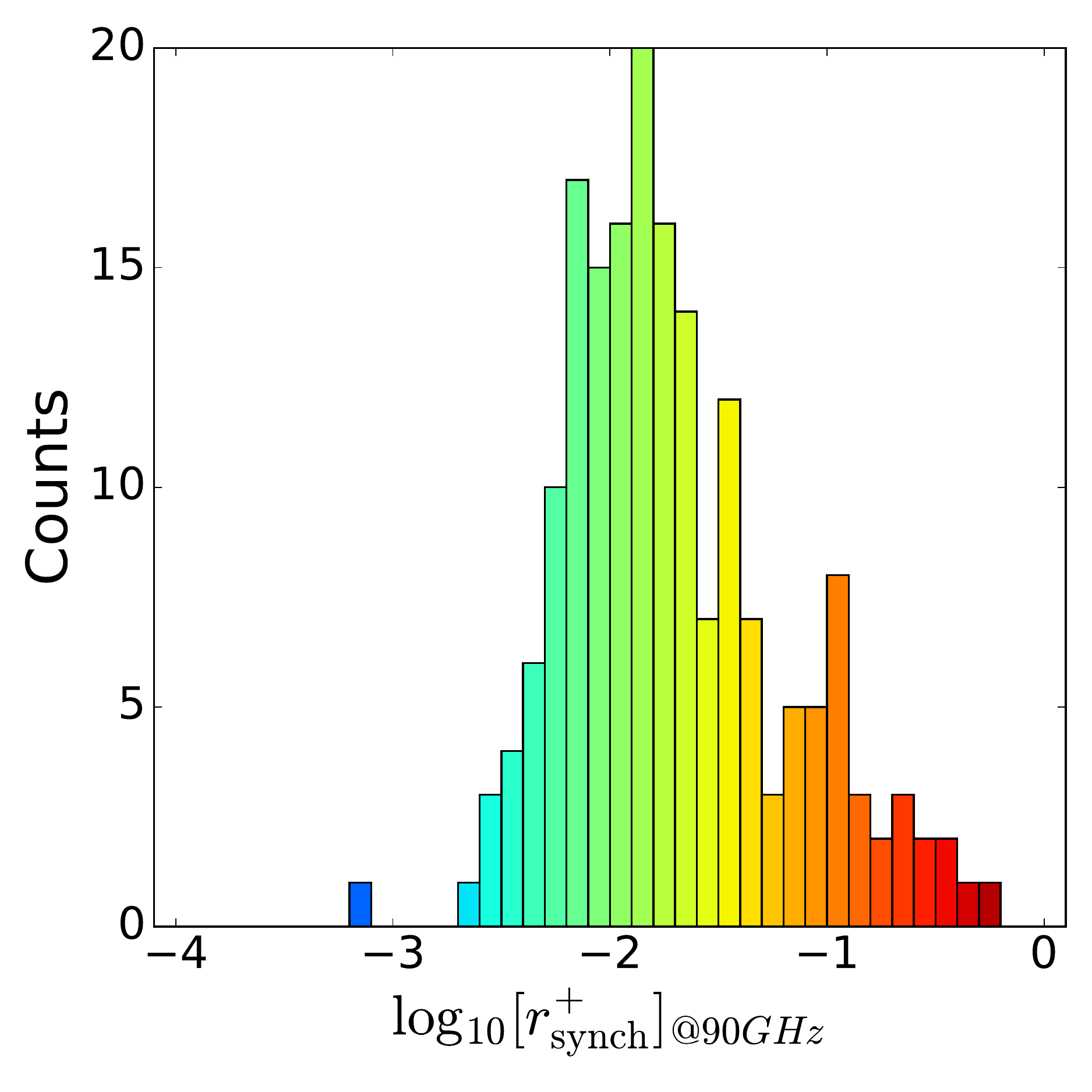}
\includegraphics[width=6cm]{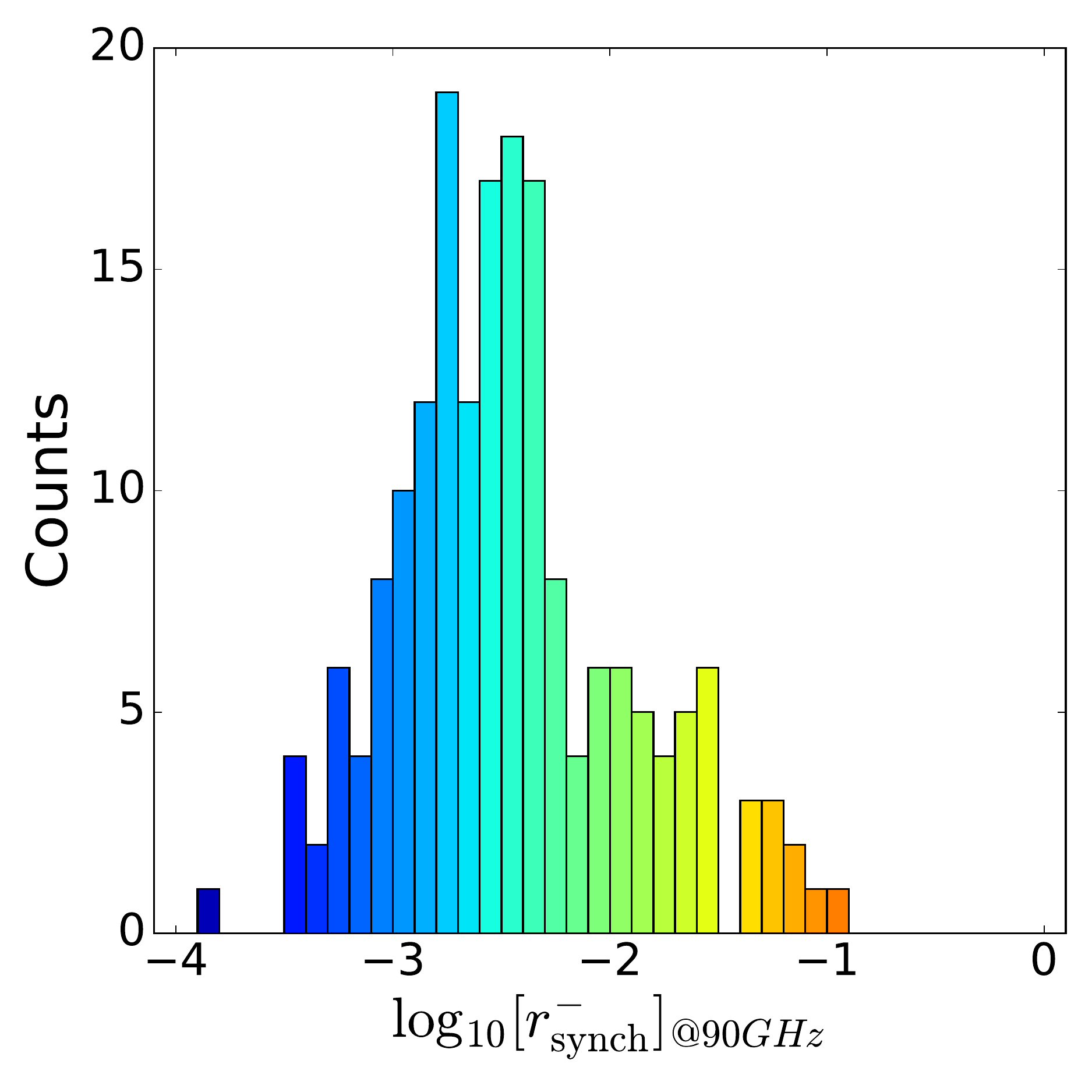}
\caption{Synchrotron contamination to CMB $B$-modes at 90 GHz from {\tt S-PASS} 
data, expressed in term of equivalent tensor-to-scalar ratio. In the top panels, 
the larger and smaller maps show the mean estimated value for $r_{\text{synch}}$ 
and the $1\sigma$ lower and upper limit, $r_{\text{synch}}^{-}$ and $r_{\text{synch}}^{+}$, 
respectively. The lower panels show the histograms of the corresponding maps.}
\label{r_synch}
\end{figure*}

As a final aspect of our analysis, we use the {\tt S-PASS} and {\tt Planck} data to estimate the level of contamination from foregrounds to the observation of primordial CMB $B$-modes. We focus on the degree angular scales, around $\ell\sim80$, where the recombination bump of the CMB $B$-modes peaks.\par 

We estimate the amplitude of foreground radiation on small sky regions, with $f_{sky}\sim1\%$, with the goal of obtaining a map of the level of contamination. This approach has been already used in \citet{planck2014-XXX} to 
estimate the contamination coming from Galactic thermal dust emission. Moreover, the analysis we present here represents an update of the work described in \citet{K16}, where the synchrotron and thermal dust signals at the 
degree scales were estimated using {\tt Planck} and {\tt WMAP} data. 

\subsection{Synchrotron contamination to primordial $B$-modes}
\label{Sec:synch_contamination}
As a first step we analyze {\tt S-PASS} data alone, with the goal of obtaining a map of the level of contamination coming only from synchrotron radiation to primordial $B$-modes. As mentioned, in this part of our work, we focus 
on small regions of the sky located at intermediate and high Galactic latitudes ($|b|>20^{\circ}$). We construct a set of 184 circular masks, each of which covers a fraction of the sky of about 1.2\%\footnote{Note that given the geometry of the portion of the sky observed by the {\tt S-PASS} survey, and the latitude cut we consider, some of the sky patches may not be circular and cover a smaller sky fraction. Nevertheless, all the 184 regions we include in the analysis have $f_{sky}>1\%$ (before applying apodization)}. The procedure we use to generate these masks is described in Section \ref{Section:mask}.\par

On each mask we compute the $B$-mode auto angular power spectrum of {\tt S-PASS} maps, using the {\tt Xpol} algorithm. Spectra are computed for multipoles ranging from $\ell=40$ to $\ell=140$, considering 5 bins with 
$\Delta\ell=20$.\par

Once we have our set of 184 $B$-mode spectra, we fit each of them considering a power law to describe the power spectrum behavior, i.e. $C_{\ell}=A^s_{\ell=80}\ell^{\,\alpha}$. We use the value $\alpha=-3$, which represents 
the average of the values we find by performing the fit on {\tt S-PASS} power spectra on the set of iso-latitude masks (see Section \ref{Section:spass_specta_fit} and Table \ref{fit_results}), and fit only for the spectra amplitude. With this procedure we obtain an estimate of the amplitude of $A^s_{\ell=80}$ of synchrotron $B$-mode spectrum at 2.3 GHz and at $\ell=80$, in each region. The error on this amplitude takes into account both the one coming from the power law fit as well as the uncertainty on {\tt S-PASS} photometric calibration that we consider to be at 5\% level on maps. \par

We assess the level of contamination coming from synchrotron to CMB $B$-modes at 90 GHz, one of the typical channels at which CMB experiments observe the sky and close to the minimum of foreground emission. To do so, we extrapolate the recovered amplitudes $A^s_{\ell=80}$, considering a power law SED, with $\beta_s=-3.22\pm0.08$. We divide these amplitudes extrapolated at 90 GHz, by the amplitude of CMB $B$-modes at $\ell=80$, considering the best {\tt Planck} 2015 $\Lambda$CDM model with tensor-to-scalar ratio $r=1$ ($A^{CMB}_{\ell=80} = 7.54\times10^{-2}\, \mu K^2$). In this way we obtain an estimate of the parameter, hereafter called $r_{\text{synch}}$, whose value describes the contamination to primordial CMB $B$-modes in term of equivalent tensor-to-scalar ratio.\par

Figure \ref{r_synch} reports the map, together with the corresponding histogram, of the value of $r_{\text{synch}}$ in the pixels of a map at $N_{side}=8$ on which each of the 
considered circular sky regions is centered. We also report the $1\sigma$ upper and lower limits on the value of $r_{\text{synch}}$, computed by taking into account both the error on $A^s_{\ell=80}$ and the uncertainty on $\beta_s$. The Figure shows new remarkable features with respect to what has been found in previous works considering satellite data \citep{K16}. There, only peaks in the synchrotron emission were detectable above noise threshold, while here the regions with lowest emission can also be measured. Results show that, even in the cleanest region of the sky, the synchrotron contamination is at the level of $r_{\text{synch}}=10^{-3}$ at 90 GHz.

\begin{figure*}[!t]
\centering
\includegraphics[width=\textwidth]{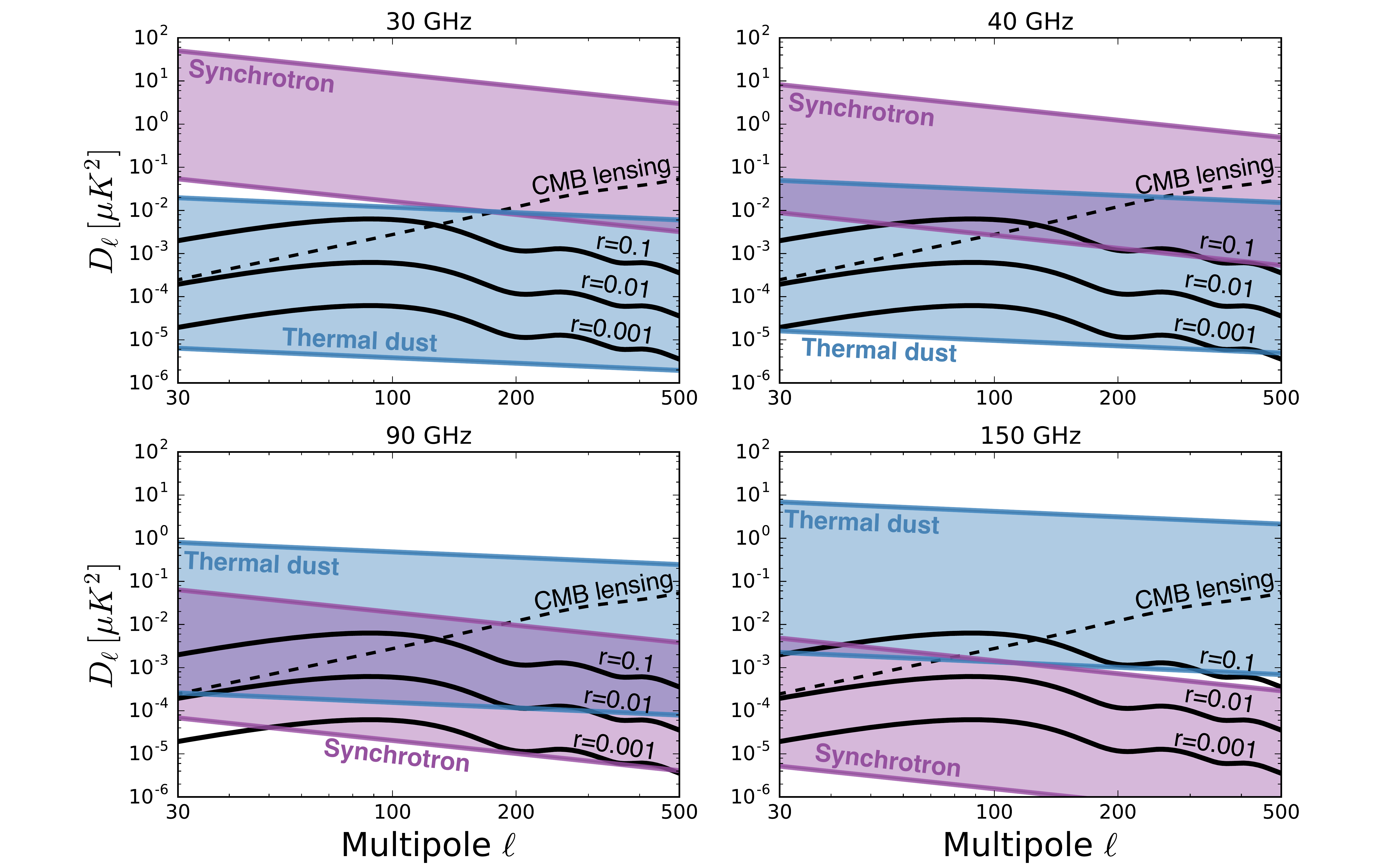}
\caption{Synchrotron (violet areas) and thermal dust (cyan areas) amplitude evaluated in 184 sky regions with $f_{sky}\simeq1\%$
extrapolated at the indicated frequencies and compared with cosmological $B$-modes with given values of $r$ 
(solid black lines, dashed for cosmological lensing again as predicted by {\tt Planck} 2015 best fit 
$\Lambda$CDM cosmologies).}
\label{r_synch_dust_spectra}
\end{figure*}

\begin{figure*}[!t]
\centering
\includegraphics[width=\textwidth]{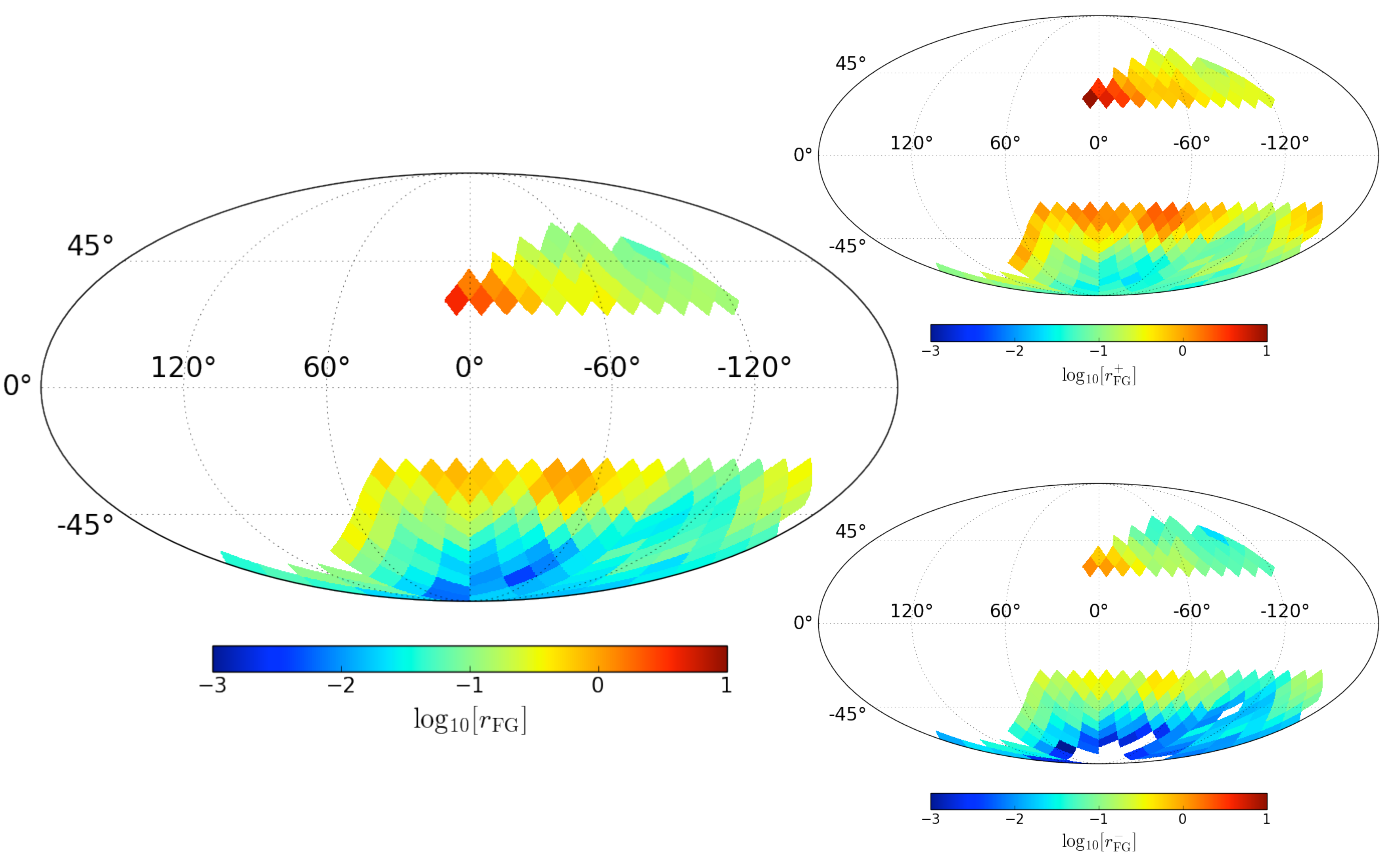}\\
\includegraphics[width=6cm]{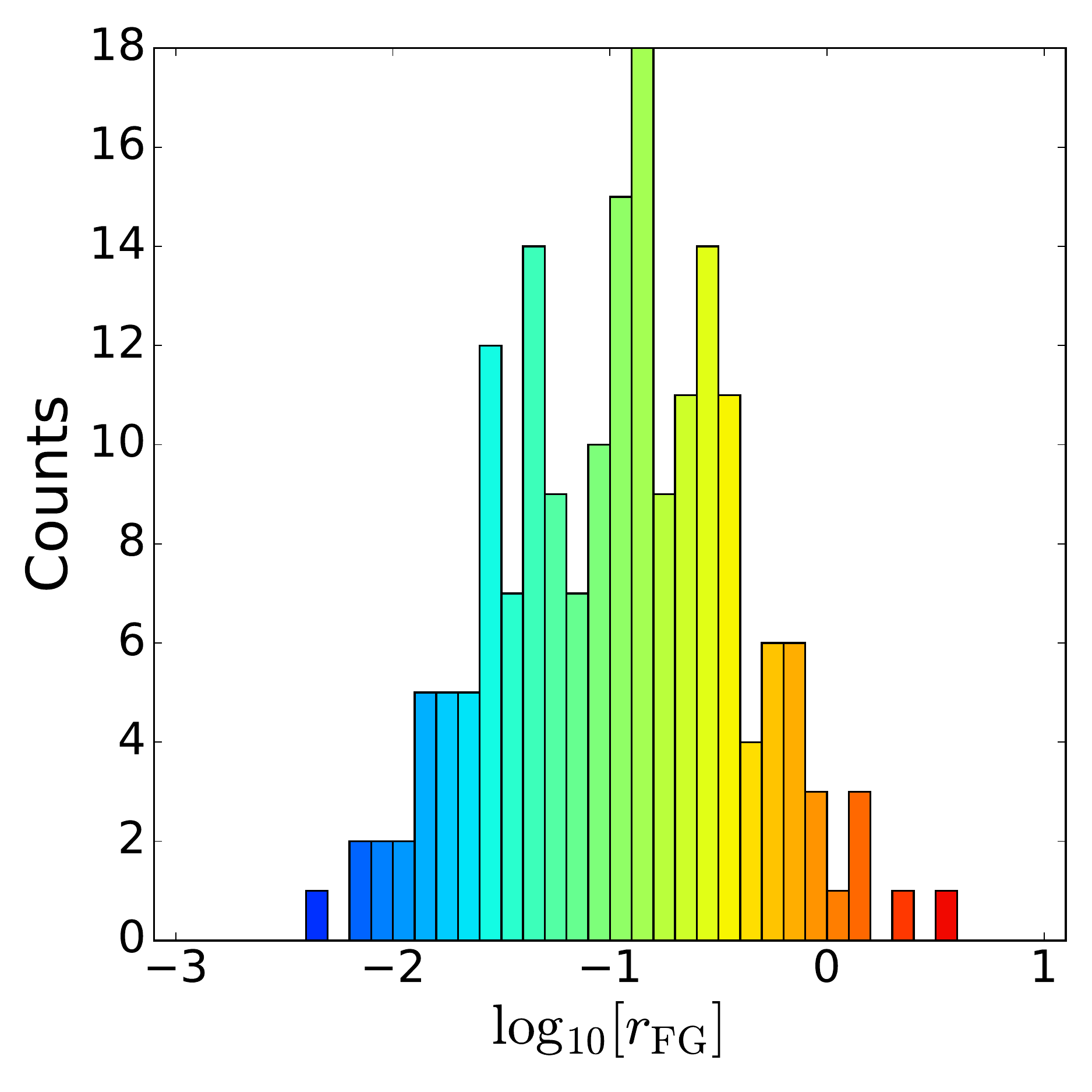}
\includegraphics[width=6cm]{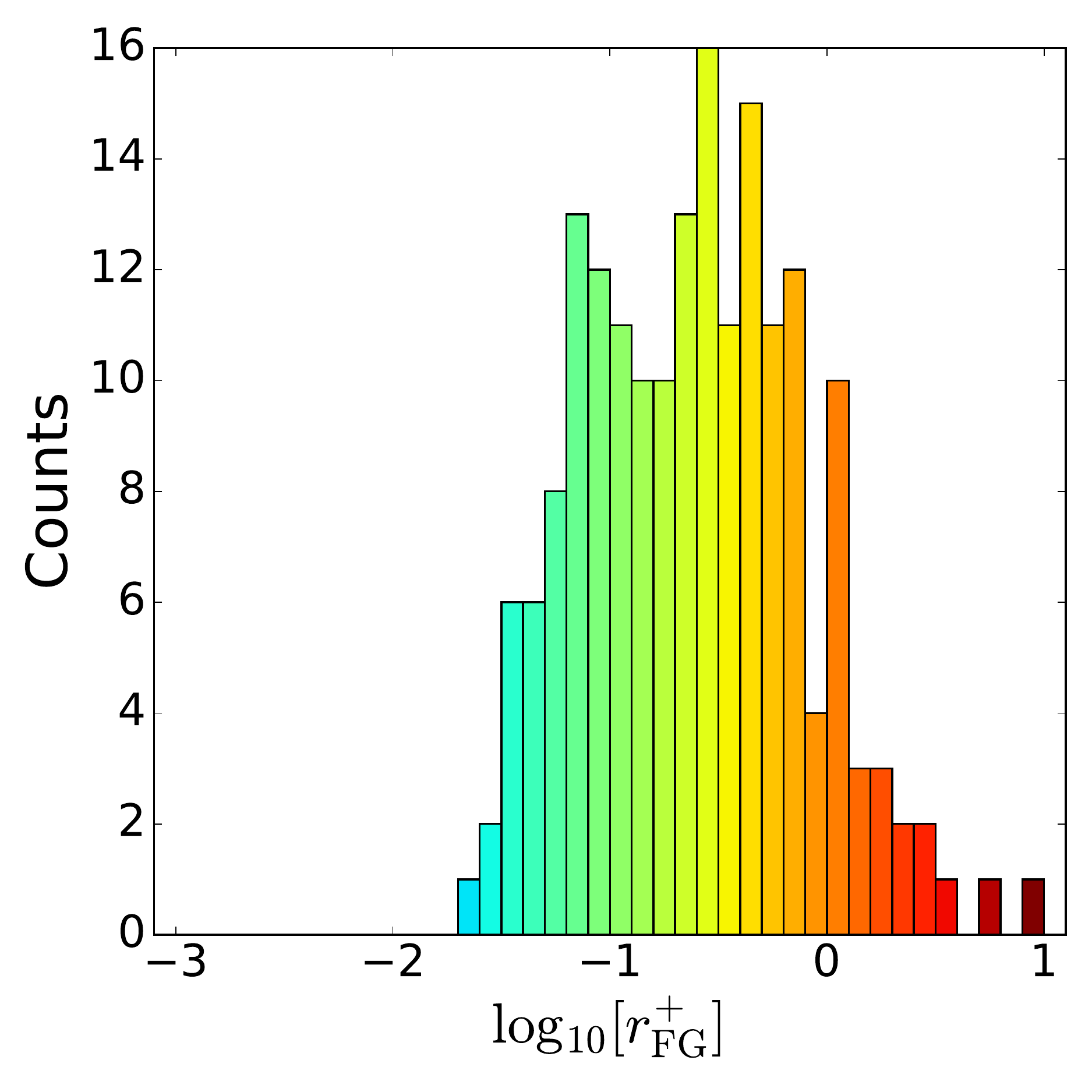}
\includegraphics[width=6cm]{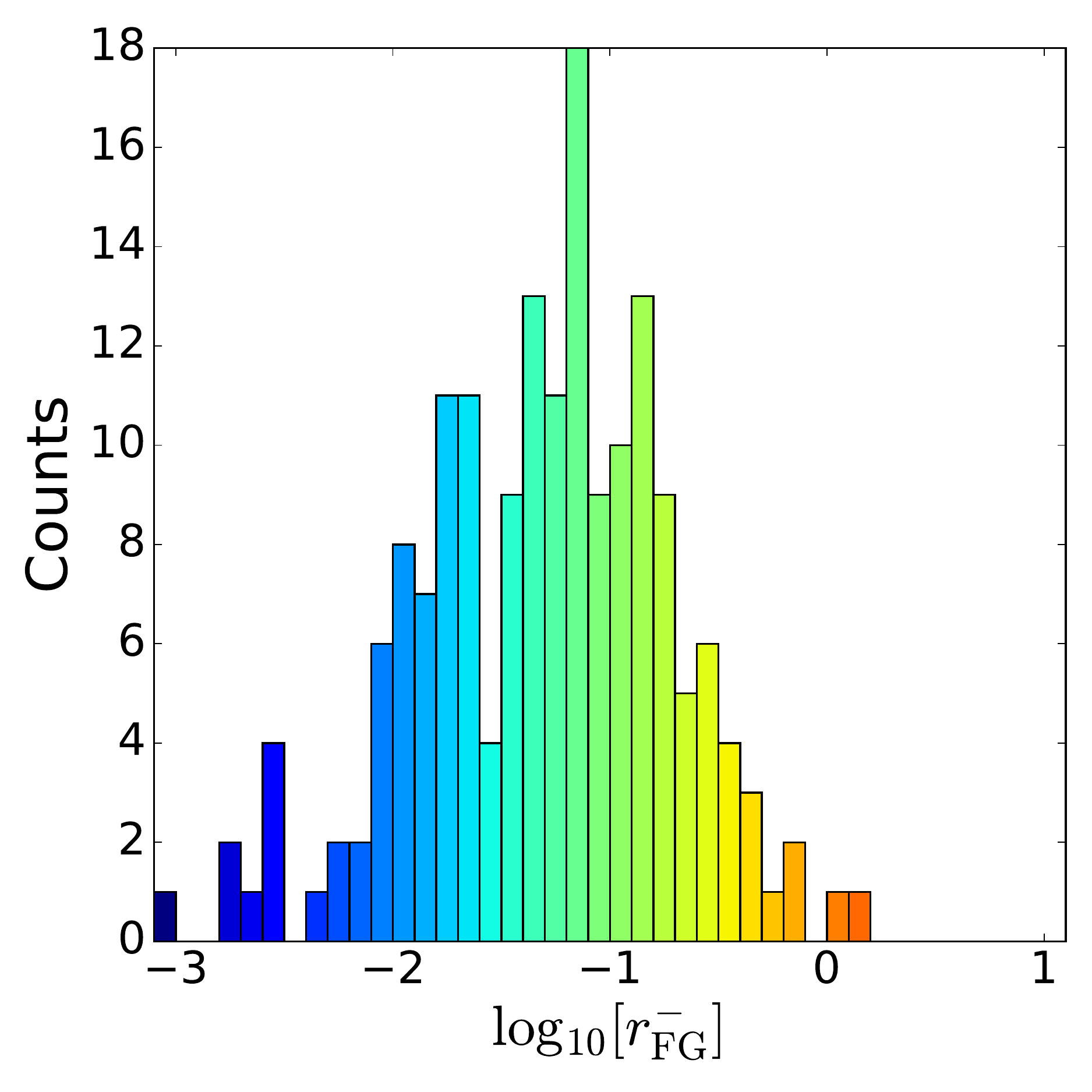}
\caption{Sky distribution of the minimum foreground emission at $\ell =80$ expressed in 
units of the cosmological tensor-to-scalar ratio, $r_{\text{FG}}$, from {\tt S-PASS} and {\tt Planck}-353 data. Larger and smaller maps show the mean estimated value for $r_{\text{FG}}$ and the $1\sigma$ lower and upper limit, $r_{\text{FG}}^{-}$ and $r_{\text{FG}}^{+}$, respectively; the white spots close to the southern Galactic pole for $r_{\text{FG}}^{-}$ represent locations where the latter is negative. The lower panels show the histograms of the corresponding maps.}
\label{r_fg_minimum}
\end{figure*}

\begin{figure*}[!t]
\centering
\includegraphics[width=9cm]{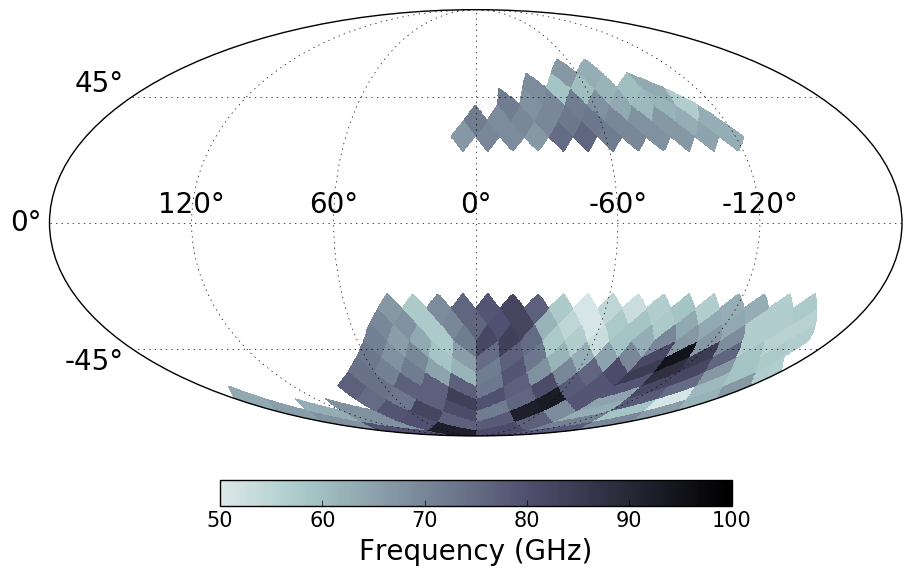}
\includegraphics[width=5cm]{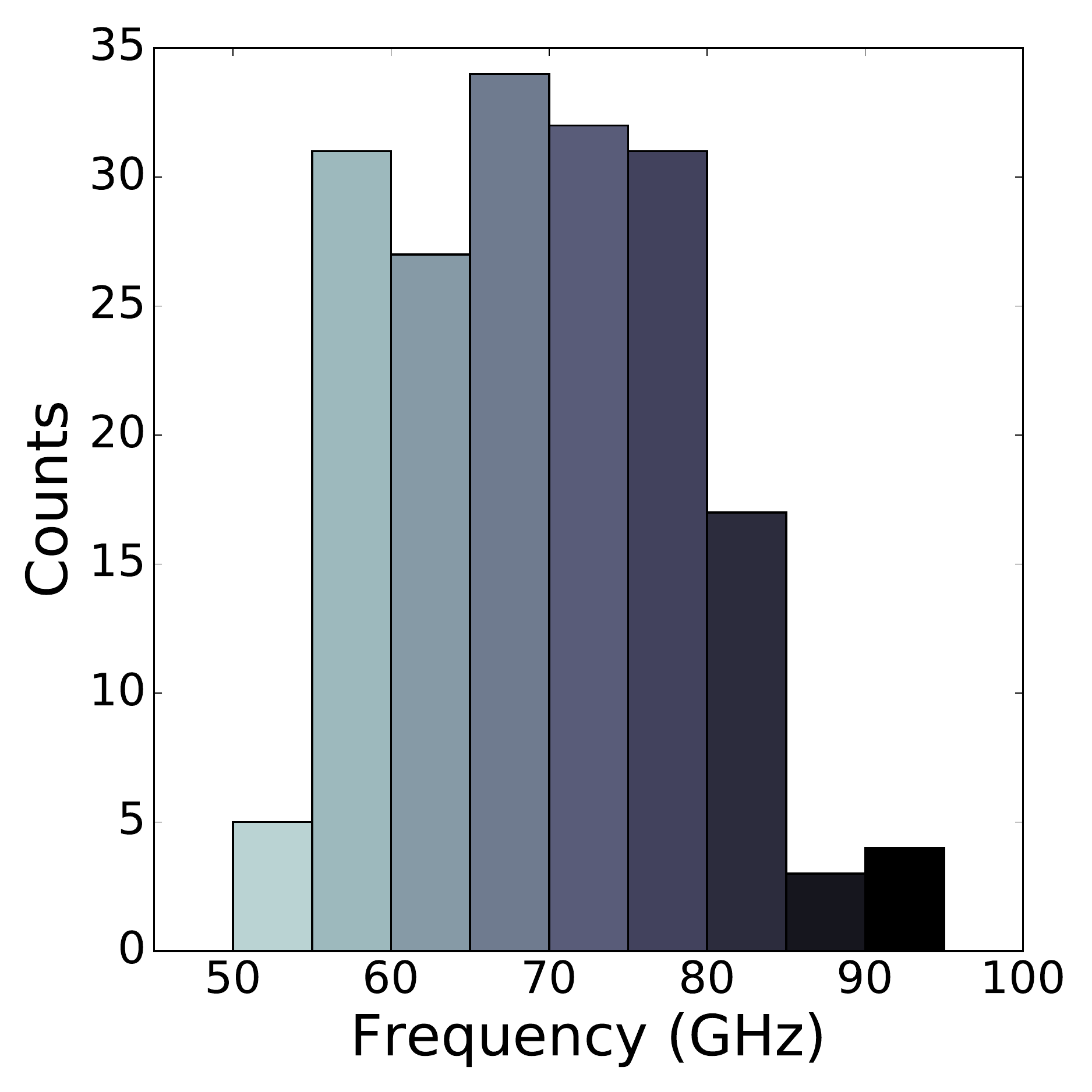}
\caption{Sky distribution (left panel) and histogram (right panel) of the frequency corresponding to $r_{\text{FG}}$ in Fig. \ref{r_fg_minimum}.}
\label{r_fg_minimum_frequency}
\end{figure*}

\subsection{Foreground minimum}

Finally, we compare the {\tt S-PASS} levels of synchrotron measured in this work with those from dust, to assess the relevance and location in frequency of the foreground minimum with respect to cosmologically interesting values of $r$.\par 

In Figure \ref{r_synch_dust_spectra} we report the synchrotron $B$-mode power spectra inferred in this work from {\tt S-PASS} data, evaluated as described in Section \ref{Sec:synch_contamination} on the 184 sky regions outlined above. For the spectral shape we use the usual power law with $\beta_s=-3.22$, to extrapolate the synchrotron amplitude at the frequencies of 30, 40, 90 and 150 GHz. These spectra are compared with those for the thermal dust emission, with amplitude evaluated by fitting the $B$-modes power spectra. We compute these spectra by cross correlating the {\tt Planck} 353 GHz half-mission maps on the same circular sky regions, with a power law with $\alpha_d=-2.42$ \citep{planck2014-XXX}, on the multipole interval $40\leq\ell\leq140$. The extrapolation in frequency for thermal dust emission is done considering a modified black body SED with $\beta_d=1.59$ and $T_d=19.6$ K \citep{planck2014-XXII}.\par

The amplitude of the two signals, in the different sky regions, are compared, at different frequencies, with cosmologically interesting levels of $r$ as well as lensing, again as predicted by the {\tt Planck} 2015 best fit $\Lambda$CDM cosmology. Results shows that there is no region of the sky, nor frequency, where the foreground level lies below the CMB power $B$-modes power spectrum with $r\simeq10^{-3}$. \par

In Figure \ref{r_fg_minimum} we show the sky distribution of the minimum of foreground emission. This is obtained by summing the amplitude of synchrotron and thermal dust $B$-modes at $\ell=80$ extrapolated at the different frequencies, using the SEDs described above, and looking for the frequency at which this sum reaches its minimum. By computing the simple sum of the two kinds of emission, here we are neglecting the correlation between the two. Levels of the minimum emission show $r_{\text{FG}}\simeq10^{-2}$ at high latitudes, while approaching unity at the lowest latitudes considered. As in Figure \ref{r_synch}, uncertainties take into account the contribution of the noise and the extrapolation in frequencies. The best location looks to be the south Galactic cap, which is thus the most opportune place where to observe for experiments aiming at the lowest possible $r$ value.\par 

In Figure \ref{r_fg_minimum_frequency}, we show the sky distribution and histogram of the frequency $\nu_{r_{\text{FG}}}$ corresponding to the $r_{\text{FG}}$ shown in Fig. \ref{r_fg_minimum}. The range of values is concentrated mostly between 60 and 90 GHz, confirming what previously inferred with a much poorer statistics \citep{K16}. The frequency distribution appears to be rather flat, when we consider all the analyzed sky patches, while, if we restrict only to the cleanest regions (with $r_{\text{FG}}<0.03$) the foreground minimum is preferably reached at frequencies around 80 GHz. 


\section{Discussion and conclusion}
\label{Section:Conc}

In this work, we have analyzed the southern sky linear polarization at 2.3 GHz as observed by the {\tt S-PASS} survey. The forthcoming {\tt S-PASS} survey paper describes in detail the data reduction and production of sky 
maps which represent the starting point of the present analysis. These measurements represent an important supplement to the observations of the synchrotron radiation carried out by the {\tt WMAP} and {\tt Planck} satellites, at higher and more interesting frequencies for CMB purposes, but with poorer angular resolution, and lower signal-to-noise ratio. \par 

The main target of our analysis has been the study of the properties of the diffuse Galactic polarized synchrotron. We have measured the angular distribution of the {\tt S-PASS} signal at increasingly high Galactic latitudes, computing the auto power spectra of the polarization maps. At large angular scale ($\ell\lesssim200$), the $C_{\ell}$ signal can be described, in first approximation, as a power law $\propto\ell^{\alpha}$. The recovered value of the index $\alpha$ changes significantly with respect to the considered sky regions. At Galactic latitudes $|b|> 30^{\circ}$ the mean value is $\alpha\simeq-3.15$, with steeper spectra for $E$ than for $B$-modes. At Galactic latitude including also portion of the sky closer to the Galactic plane, the signal is contaminated by Faraday rotation effects, causing polarization angle modulation, and inducing a mixing between the polarization states and flatter power spectra. Synchrotron signal also shows an asymmetry between $B$ and $E$-modes emission, with a $B$-to-$E$ ratio $\sim0.5$ for $|b|> 35^{\circ}$.\par

We have combined {\tt S-PASS} data with those from {\tt WMAP} and {\tt Planck} satellites (at the frequencies of 23, 33 and 28.4 GHz) to obtain information about the synchrotron SED. We have performed this analysis in either harmonic or pixel space. \par

In the first case we have computed power spectra of the polarization maps of the different instruments, and have fit the amplitude of those spectra, as a function of frequency, separately for different multipole bins and sky masks. The recovered mean value for the synchrotron spectral index is $\beta_s=-3.22\pm0.08$, which appears to be constant in the entire multipole range we have studied (six bins in the interval $20\leq\ell<140$) and sky regions. Also, we have not observed any significant difference in the spectral behavior of $E$ and $B$-modes. Although we have fit the data in an unprecedented range of frequency (from 2.3 to 33 GHz), the value we find for the $\beta_s$ index is in agreement with what measured from {\tt Planck} and {\tt WMAP} at higher frequencies. We also have fit the data with a model allowing a curvature for the synchrotron spectral index, that we find to be compatible with zero. We however emphasize the need of more data at intermediate frequencies to better constrain this parameter. We have seen evidence of de-correlation at mid latitudes, with a deficit of the power measured in the cross-spectra between {\tt S-PASS} and {\tt WMAP}/{\tt Planck}. This de-correlation is strong close to the Galactic plane, indicating that, in these regions, Faraday rotation in {\tt S-PASS} data plays an important role. On the other hand, the effect is not statistically significant at high latitudes.\par

We have also analyzed the synchrotron SED in pixel space, constructing a map of the synchrotron spectral index. We have used again the {\tt S-PASS}/{\tt WMAP}/{\tt Planck} data, covering the frequency range 2.3-33 GHz. To obtain the $\beta_s$ map we have fit the data in total polarization at $2^{\circ}$ angular resolution. The distribution of $\beta_s$ on the map peaks around the value -3.2, in agreement with what we have found from the SED analysis in the harmonic domain. The angular power spectrum of this map can be approximated with a power law, with $C_{\ell}^{\beta_s\beta_s}\propto\ell^{\gamma}$ and $\gamma=-2.6\pm0.2$. We stress here that this represents the first available $\beta_s$ map, and associated power spectrum, obtained using only polarization data, and therefore will be useful to build realistic simulations of the synchrotron signal as input for assessing the performance of component separation algorithms.\par

By cross correlating {\tt S-PASS} data with the high frequency polarization maps coming from the {\tt Planck}-HFI 353 GHz channel, we have measured the spatial correlation between synchrotron and thermal dust signals. We find evidence for positive correlation, reaching about $40\%$ on large angular scales, gently decaying to smaller values on smaller scales, depending on the latitudes considered. These results confirm the previous ones obtained from joint {\tt Planck} and {\tt WMAP} analysis, showing that the correlation among the two different kind of signals, on the large angular scale, persists over a wide interval of frequencies.\par

Finally, by taking advantage of the high signal-to-noise regime of the {\tt S-PASS} survey, and our measured synchrotron SED, we have refined the estimation of the synchrotron contamination to CMB $B$-modes. We have constructed a map reporting the level of contamination at the degree angular scale ($\ell=80$) and at 90 GHz in 184 sky patch, each of which covers a sky fraction of about 1\%. Results show that the minimum contamination is at the level of an equivalent tensor-to-scalar ratio $r_{synch}\simeq10^{-3}$. We have also combined synchrotron and thermal dust information, getting a map of the estimated total level of contamination coming from foregrounds. Our findings, confirm, once again, that there is no region of the sky (among the sky portion covered by the {\tt S-PASS} survey) nor frequency where the foreground amplitude (at the degree angular scales) lie below a CMB $B$-mode signal with $r\simeq10^{-3}$, and that therefore frequency channel monitoring foreground emission, on both low and high frequency, are mandatory for all the experiments aiming at observing the primordial GWs signal at this level.\par

We conclude stressing the importance of radio survey as tracer for synchrotron radiation. Given the brightness of the emission at these frequencies they can reach high signal-to-noise ratio, giving therefore fundamental information about the characteristics of the signal itself, precious to build realistic sky models. Nevertheless, new data at intermediate frequencies are needed, in order to better characterize the synchrotron SED, better constrain the curvature of the spectral index, and possibly identify the presence of frequency de-correlation and its physical nature. Additional data will also be of fundamental importance to understand if radio observations can also be used as complementary data for CMB experiments to perform component separation and to isolate the cosmological signal.


   \begin{acknowledgements}
The Parkes Radio Telescope is part of the Australia Telescope National Facility, which is funded by the Commonwealth of Australia for operation as a National Facility managed by CSIRO. This work has been carried out in the framework of the S-band Polarization All Sky Survey ({\tt S-PASS}) collaboration. This research was supported by the RADIOFOREGROUNDS project, funded by the European Commissions H2020 Research Infrastructures under the Grant Agreement 687312. CB acknowledges support by the INDARK INFN Initiative. We acknowledge support from the ASI-COSMOS Network (cosmosnet.it).
\end{acknowledgements}

\bibliographystyle{aa} 
\bibliography{ms} 

\begin{thebibliography}{54}
\expandafter\ifx\csname natexlab\endcsname\relax\def\natexlab#1{#1}\fi

\bibitem[{{Abazajian} {et~al.}(2016){Abazajian}, {Adshead}, {Ahmed}, {Allen},
  {Alonso}, {Arnold}, {Baccigalupi}, {Bartlett}, {Battaglia}, {Benson},
  {Bischoff}, {Borrill}, {Buza}, {Calabrese}, {Caldwell}, {Carlstrom}, {Chang},
  {Crawford}, {Cyr-Racine}, {De Bernardis}, {de Haan}, {di Serego Alighieri},
  {Dunkley}, {Dvorkin}, {Errard}, {Fabbian}, {Feeney}, {Ferraro}, {Filippini},
  {Flauger}, {Fuller}, {Gluscevic}, {Green}, {Grin}, {Grohs}, {Henning},
  {Hill}, {Hlozek}, {Holder}, {Holzapfel}, {Hu}, {Huffenberger}, {Keskitalo},
  {Knox}, {Kosowsky}, {Kovac}, {Kovetz}, {Kuo}, {Kusaka}, {Le Jeune}, {Lee},
  {Lilley}, {Loverde}, {Madhavacheril}, {Mantz}, {Marsh}, {McMahon},
  {Meerburg}, {Meyers}, {Miller}, {Munoz}, {Nguyen}, {Niemack}, {Peloso},
  {Peloton}, {Pogosian}, {Pryke}, {Raveri}, {Reichardt}, {Rocha}, {Rotti},
  {Schaan}, {Schmittfull}, {Scott}, {Sehgal}, {Shandera}, {Sherwin}, {Smith},
  {Sorbo}, {Starkman}, {Story}, {van Engelen}, {Vieira}, {Watson}, {Whitehorn},
  \& {Kimmy Wu}}]{2016arXiv161002743A}
{Abazajian}, K.~N., {Adshead}, P., {Ahmed}, Z., {et~al.} 2016, ArXiv e-prints:
  1610.02743

\bibitem[{{Aiola} {et~al.}(2012){Aiola}, {Amico}, {Battaglia}, {Battistelli},
  {Ba{\'o}}, {de Bernardis}, {Bersanelli}, {Boscaleri}, {Cavaliere},
  {Coppolecchia}, {Cruciani}, {Cuttaia}, {D'Addabbo}, {D'Alessandro}, {De
  Gregori}, {Del Torto}, {De Petris}, {Fiorineschi}, {Franceschet},
  {Franceschi}, {Gervasi}, {Goldie}, {Gregorio}, {Haynes}, {Krachmalnicoff},
  {Lamagna}, {Maffei}, {Maino}, {Masi}, {Mennella}, {Morgante}, {Nati}, {Ng},
  {Pagano}, {Passerini}, {Peverini}, {Piacentini}, {Piccirillo}, {Pisano},
  {Ricciardi}, {Rissone}, {Romeo}, {Salatino}, {Sandri}, {Schillaci},
  {Stringhetti}, {Tartari}, {Tascone}, {Terenzi}, {Tomasi}, {Tommasi}, {Villa},
  {Virone}, {Withington}, {Zacchei}, \& {Zannoni}}]{2012SPIE.8446E..7AA}
{Aiola}, S., {Amico}, G., {Battaglia}, P., {et~al.} 2012, in \procspie, Vol.
  8446, Ground-based and Airborne Instrumentation for Astronomy IV, 84467A

\bibitem[{{Baccigalupi} {et~al.}(2001){Baccigalupi}, {Burigana}, {Perrotta},
  {De Zotti}, {La Porta}, {Maino}, {Maris}, \&
  {Paladini}}]{2001A&A...372....8B}
{Baccigalupi}, C., {Burigana}, C., {Perrotta}, F., {et~al.} 2001, \aap, 372, 8

\bibitem[{{Bennett} {et~al.}(2013){Bennett}, {Larson}, {Weiland}, {Jarosik},
  {Hinshaw}, {Odegard}, {Smith}, {Hill}, {Gold}, {Halpern}, {Komatsu}, {Nolta},
  {Page}, {Spergel}, {Wollack}, {Dunkley}, {Kogut}, {Limon}, {Meyer}, {Tucker},
  \& {Wright}}]{2013ApJS..208...20B}
{Bennett}, C.~L., {Larson}, D., {Weiland}, J.~L., {et~al.} 2013, \apjs, 208, 20

\bibitem[{{BICEP2 and Keck Array Collaborations}(2016)}]{2016PhRvL.116c1302B}
{BICEP2 and Keck Array Collaborations}. 2016, Physical Review Letters, 116,
  031302

\bibitem[{{BICEP2/Keck and Planck Collaborations}(2015)}]{B2P}
{BICEP2/Keck and Planck Collaborations}. 2015, Physical Review Letters, 114,
  101301

\bibitem[{{Carretti}(2010)}]{2010ASPC..438..276C}
{Carretti}, E. 2010, in Astronomical Society of the Pacific Conference Series,
  Vol. 438, 276

\bibitem[{{Carretti} {et~al.}(2013){Carretti}, {Crocker}, {Staveley-Smith},
  {Haverkorn}, {Purcell}, {Gaensler}, {Bernardi}, {Kesteven}, \&
  {Poppi}}]{2013Natur.493...66C}
{Carretti}, E., {Crocker}, R.~M., {Staveley-Smith}, L., {et~al.} 2013, \nat,
  493, 66

\bibitem[{{Carretti} {et~al.}(2010){Carretti}, {Haverkorn}, {McConnell},
  {Bernardi}, {McClure-Griffiths}, {Cortiglioni}, \&
  {Poppi}}]{2010MNRAS.405.1670C}
{Carretti}, E., {Haverkorn}, M., {McConnell}, D., {et~al.} 2010, \mnras, 405,
  1670

\bibitem[{{{Carretti} et al.}(2018)}]{carretti_etal_in_preparation}
{{Carretti} et al.} 2018, in prep.

\bibitem[{{Choi} \& {Page}(2015)}]{2015JCAP...12..020C}
{Choi}, S.~K. \& {Page}, L.~A. 2015, \jcap, 12, 020

\bibitem[{{Delabrouille} {et~al.}(2017){Delabrouille}, {de Bernardis},
  {Bouchet}, {Ach{\'u}carro}, {Ade}, {Allison}, {Arroja}, {Artal}, {Ashdown},
  {Baccigalupi}, {Ballardini}, {Banday}, {Banerji}, {Barbosa}, {Bartlett},
  {Bartolo}, {Basak}, {Baselmans}, {Basu}, {Battistelli}, {Battye}, {Baumann},
  {Beno{\^\i}t}, {Bersanelli}, {Bideaud}, {Biesiada}, {Bilicki}, {Bonaldi},
  {Bonato}, {Borrill}, {Boulanger}, {Brinckmann}, {Brown}, {Bucher},
  {Burigana}, {Buzzelli}, {Cabass}, {Cai}, {Calvo}, {Caputo}, {Carvalho},
  {Casas}, {Castellano}, {Catalano}, {Challinor}, {Charles}, {Chluba},
  {Clements}, {Clesse}, {Colafrancesco}, {Colantoni}, {Contreras},
  {Coppolecchia}, {Crook}, {D'Alessandro}, {D'Amico}, {da Silva}, {de Avillez},
  {de Gasperis}, {De Petris}, {de Zotti}, {Danese}, {D{\'e}sert}, {Desjacques},
  {Di Valentino}, {Dickinson}, {Diego}, {Doyle}, {Durrer}, {Dvorkin},
  {Eriksen}, {Errard}, {Feeney}, {Fern{\'a}ndez-Cobos}, {Finelli},
  {Forastieri}, {Franceschet}, {Fuskeland}, {Galli}, {G{\'e}nova-Santos},
  {Gerbino}, {Giusarma}, {Gomez}, {Gonz{\'a}lez-Nuevo}, {Grandis},
  {Greenslade}, {Goupy}, {Hagstotz}, {Hanany}, {Handley},
  {Henrot-Versill{\'e}}, {Hern{\'a}ndez-Monteagudo}, {Hervias-Caimapo},
  {Hills}, {Hindmarsh}, {Hivon}, {Hoang}, {Hooper}, {Hu}, {Keih{\"a}nen},
  {Keskitalo}, {Kiiveri}, {Kisner}, {Kitching}, {Kunz}, {Kurki-Suonio},
  {Lagache}, {Lamagna}, {Lapi}, {Lasenby}, {Lattanzi}, {Le Brun},
  {Lesgourgues}, {Liguori}, {Lindholm}, {Lizarraga}, {Luzzi},
  {Mac{\`\i}as-P{\'e}rez}, {Maffei}, {Mandolesi}, {Martin},
  {Martinez-Gonzalez}, {Martins}, {Masi}, {Massardi}, {Matarrese}, {Mazzotta},
  {McCarthy}, {Melchiorri}, {Melin}, {Mennella}, {Mohr}, {Molinari},
  {Monfardini}, {Montier}, {Natoli}, {Negrello}, {Notari}, {Noviello},
  {Oppizzi}, {O'Sullivan}, {Pagano}, {Paiella}, {Pajer}, {Paoletti},
  {Paradiso}, {Partridge}, {Patanchon}, {Patil}, {Perdereau}, {Piacentini},
  {Piat}, {Pisano}, {Polastri}, {Polenta}, {Pollo}, {Ponthieu}, {Poulin},
  {Pr{\^e}le}, {Quartin}, {Ravenni}, {Remazeilles}, {Renzi}, {Ringeval},
  {Roest}, {Roman}, {Roukema}, {Rubino-Martin}, {Salvati}, {Scott}, {Serjeant},
  {Signorelli}, {Starobinsky}, {Sunyaev}, {Tan}, {Tartari}, {Tasinato},
  {Toffolatti}, {Tomasi}, {Torrado}, {Tramonte}, {Trappe}, {Triqueneaux},
  {Tristram}, {Trombetti}, {Tucci}, {Tucker}, {Urrestilla}, {V{\"a}liviita},
  {Van de Weygaert}, {Van Tent}, {Vennin}, {Verde}, {Vermeulen}, {Vielva},
  {Vittorio}, {Voisin}, {Wallis}, {Wandelt}, {Wehus}, {Weller}, {Young},
  {Zannoni}, \& {for the CORE collaboration}}]{2017arXiv170604516D}
{Delabrouille}, J., {de Bernardis}, P., {Bouchet}, F.~R., {et~al.} 2017, ArXiv
  e-prints: 1706.04516

\bibitem[{{Errard} {et~al.}(2016){Errard}, {Feeney}, {Peiris}, \&
  {Jaffe}}]{2016JCAP...03..052E}
{Errard}, J., {Feeney}, S.~M., {Peiris}, H.~V., \& {Jaffe}, A.~H. 2016, \jcap,
  3, 052

\bibitem[{{Essinger-Hileman} {et~al.}(2014){Essinger-Hileman}, {Ali}, {Amiri},
  {Appel}, {Araujo}, {Bennett}, {Boone}, {Chan}, {Cho}, {Chuss}, {Colazo},
  {Crowe}, {Denis}, {D{\"u}nner}, {Eimer}, {Gothe}, {Halpern}, {Harrington},
  {Hilton}, {Hinshaw}, {Huang}, {Irwin}, {Jones}, {Karakla}, {Kogut}, {Larson},
  {Limon}, {Lowry}, {Marriage}, {Mehrle}, {Miller}, {Miller}, {Moseley},
  {Novak}, {Reintsema}, {Rostem}, {Stevenson}, {Towner}, {U-Yen}, {Wagner},
  {Watts}, {Wollack}, {Xu}, \& {Zeng}}]{CLASS}
{Essinger-Hileman}, T., {Ali}, A., {Amiri}, M., {et~al.} 2014, in Society of
  Photo-Optical Instrumentation Engineers (SPIE) Conference Series, Vol. 9153,
  Society of Photo-Optical Instrumentation Engineers (SPIE) Conference Series,
  1

\bibitem[{{Fuskeland} {et~al.}(2014){Fuskeland}, {Wehus}, {Eriksen}, \&
  {N{\ae}ss}}]{Fuskeland14}
{Fuskeland}, U., {Wehus}, I.~K., {Eriksen}, H.~K., \& {N{\ae}ss}, S.~K. 2014,
  ApJ, 790, 104

\bibitem[{{G{\'e}nova-Santos} {et~al.}(2017){G{\'e}nova-Santos},
  {Rubi{\~n}o-Mart{\'{\i}}n}, {Pel{\'a}ez-Santos}, {Poidevin}, {Rebolo},
  {Vignaga}, {Artal}, {Harper}, {Hoyland}, {Lasenby},
  {Mart{\'{\i}}nez-Gonz{\'a}lez}, {Piccirillo}, {Tramonte}, \&
  {Watson}}]{2017MNRAS.464.4107G}
{G{\'e}nova-Santos}, R., {Rubi{\~n}o-Mart{\'{\i}}n}, J.~A.,
  {Pel{\'a}ez-Santos}, A., {et~al.} 2017, \mnras, 464, 4107

\bibitem[{{G{\'o}rski} {et~al.}(2005){G{\'o}rski}, {Hivon}, {Banday},
  {Wandelt}, {Hansen}, {Reinecke}, \& {Bartelmann}}]{Gorski05}
{G{\'o}rski}, K.~M., {Hivon}, E., {Banday}, A.~J., {et~al.} 2005, \apj, 622,
  759

\bibitem[{{Hanson} {et~al.}(2013){Hanson}, {Hoover}, {Crites}, {Ade}, {Aird},
  {Austermann}, {Beall}, {Bender}, {Benson}, {Bleem}, {Bock}, {Carlstrom},
  {Chang}, {Chiang}, {Cho}, {Conley}, {Crawford}, {de Haan}, {Dobbs},
  {Everett}, {Gallicchio}, {Gao}, {George}, {Halverson}, {Harrington},
  {Henning}, {Hilton}, {Holder}, {Holzapfel}, {Hrubes}, {Huang}, {Hubmayr},
  {Irwin}, {Keisler}, {Knox}, {Lee}, {Leitch}, {Li}, {Liang}, {Luong-Van},
  {Marsden}, {McMahon}, {Mehl}, {Meyer}, {Mocanu}, {Montroy}, {Natoli},
  {Nibarger}, {Novosad}, {Padin}, {Pryke}, {Reichardt}, {Ruhl}, {Saliwanchik},
  {Sayre}, {Schaffer}, {Schulz}, {Smecher}, {Stark}, {Story}, {Tucker},
  {Vanderlinde}, {Vieira}, {Viero}, {Wang}, {Yefremenko}, {Zahn}, \&
  {Zemcov}}]{2013PhRvL.111n1301H}
{Hanson}, D., {Hoover}, S., {Crites}, A., {et~al.} 2013, Physical Review
  Letters, 111, 141301

\bibitem[{{Haslam} {et~al.}(1981){Haslam}, {Klein}, {Salter}, {Stoffel},
  {Wilson}, {Cleary}, {Cooke}, \& {Thomasson}}]{1981A&A...100..209H}
{Haslam}, C.~G.~T., {Klein}, U., {Salter}, C.~J., {et~al.} 1981, \aap, 100, 209

\bibitem[{{Hu} \& {White}(1997)}]{1997NewA....2..323H}
{Hu}, W. \& {White}, M. 1997, \na, 2, 323

\bibitem[{{Jones} {et~al.}(2018){Jones}, {Taylor}, {Aich}, {Copley}, {Chiang},
  {Davis}, {Dickinson}, {Grumitt}, {Hafez}, {Heilgendorff}, {Holler}, {Irfan},
  {Jew}, {John}, {Jonas}, {King}, {Leahy}, {Leech}, {Leitch}, {Muchovej},
  {Pearson}, {Peel}, {Readhead}, {Sievers}, {Stevenson}, \&
  {Zuntz}}]{2018arXiv180504490J}
{Jones}, M.~E., {Taylor}, A.~C., {Aich}, M., {et~al.} 2018, ArXiv e-prints:
  1805.04490 [\eprint[arXiv]{1805.04490}]

\bibitem[{{Kamionkowski} {et~al.}(1997){Kamionkowski}, {Kosowsky}, \&
  {Stebbins}}]{1997PhRvD..55.7368K}
{Kamionkowski}, M., {Kosowsky}, A., \& {Stebbins}, A. 1997, \prd, 55, 7368

\bibitem[{{Kogut}(2012)}]{2012ApJ...753..110K}
{Kogut}, A. 2012, \apj, 753, 110

\bibitem[{{Kogut} {et~al.}(2007){Kogut}, {Dunkley}, {Bennett}, {Dor{\'e}},
  {Gold}, {Halpern}, {Hinshaw}, {Jarosik}, {Komatsu}, {Nolta}, {Odegard},
  {Page}, {Spergel}, {Tucker}, {Weiland}, {Wollack}, \&
  {Wright}}]{2007ApJ...665..355K}
{Kogut}, A., {Dunkley}, J., {Bennett}, C.~L., {et~al.} 2007, \apj, 665, 355

\bibitem[{{Krachmalnicoff} {et~al.}(2016){Krachmalnicoff}, {Baccigalupi},
  {Aumont}, {Bersanelli}, \& {Mennella}}]{K16}
{Krachmalnicoff}, N., {Baccigalupi}, C., {Aumont}, J., {Bersanelli}, M., \&
  {Mennella}, A. 2016, \aap, 588, A65

\bibitem[{{Lamee} {et~al.}(2016){Lamee}, {Rudnick}, {Farnes}, {Carretti},
  {Gaensler}, {Haverkorn}, \& {Poppi}}]{2016ApJ...829....5L}
{Lamee}, M., {Rudnick}, L., {Farnes}, J.~S., {et~al.} 2016, \apj, 829, 5

\bibitem[{{Louis} {et~al.}(2017){Louis}, {Grace}, {Hasselfield}, {Lungu},
  {Maurin}, {Addison}, {Ade}, {Aiola}, {Allison}, {Amiri}, {Angile},
  {Battaglia}, {Beall}, {de Bernardis}, {Bond}, {Britton}, {Calabrese}, {Cho},
  {Choi}, {Coughlin}, {Crichton}, {Crowley}, {Datta}, {Devlin}, {Dicker},
  {Dunkley}, {D{\"u}nner}, {Ferraro}, {Fox}, {Gallardo}, {Gralla}, {Halpern},
  {Henderson}, {Hill}, {Hilton}, {Hilton}, {Hincks}, {Hlozek}, {Ho}, {Huang},
  {Hubmayr}, {Huffenberger}, {Hughes}, {Infante}, {Irwin}, {Muya Kasanda},
  {Klein}, {Koopman}, {Kosowsky}, {Li}, {Madhavacheril}, {Marriage}, {McMahon},
  {Menanteau}, {Moodley}, {Munson}, {Naess}, {Nati}, {Newburgh}, {Nibarger},
  {Niemack}, {Nolta}, {Nu{\~n}ez}, {Page}, {Pappas}, {Partridge}, {Rojas},
  {Schaan}, {Schmitt}, {Sehgal}, {Sherwin}, {Sievers}, {Simon}, {Spergel},
  {Staggs}, {Switzer}, {Thornton}, {Trac}, {Treu}, {Tucker}, {Van Engelen},
  {Ward}, \& {Wollack}}]{2017JCAP...06..031L}
{Louis}, T., {Grace}, E., {Hasselfield}, M., {et~al.} 2017, \jcap, 6, 031

\bibitem[{{Lyth} {et~al.}(2010){Lyth}, {Liddle}, \& {Ma}}]{2010PhT....63g..49L}
{Lyth}, D.~H., {Liddle}, A.~R., \& {Ma}, C.-P. 2010, Physics Today, 63, 49

\bibitem[{{Matsumura} {et~al.}(2016){Matsumura}, {Akiba}, {Arnold}, {Borrill},
  {Chendra}, {Chinone}, {Cukierman}, {de Haan}, {Dobbs}, {Dominjon},
  {Elleflot}, {Errard}, {Fujino}, {Fuke}, {Goeckner-wald}, {Halverson},
  {Harvey}, {Hasegawa}, {Hattori}, {Hattori}, {Hazumi}, {Hill}, {Hilton},
  {Holzapfel}, {Hori}, {Hubmayr}, {Ichiki}, {Inatani}, {Inoue}, {Inoue},
  {Irie}, {Irwin}, {Ishino}, {Ishitsuka}, {Jeong}, {Karatsu}, {Kashima},
  {Katayama}, {Kawano}, {Keating}, {Kibayashi}, {Kibe}, {Kida}, {Kimura},
  {Kimura}, {Kohri}, {Komatsu}, {Kuo}, {Kuromiya}, {Kusaka}, {Lee}, {Linder},
  {Matsuhara}, {Matsuoka}, {Matsuura}, {Mima}, {Mitsuda}, {Mizukami}, {Morii},
  {Morishima}, {Nagai}, {Nagasaki}, {Nagata}, {Nakajima}, {Nakamura},
  {Namikawa}, {Naruse}, {Natsume}, {Nishibori}, {Nishijo}, {Nishino}, {Nitta},
  {Noda}, {Noguchi}, {Ogawa}, {Oguri}, {Ohta}, {Otani}, {Okada}, {Okamoto},
  {Okamoto}, {Okamura}, {Rebeiz}, {Richards}, {Sakai}, {Sato}, {Sato},
  {Segawa}, {Sekiguchi}, {Sekimoto}, {Sekine}, {Seljak}, {Sherwin},
  {Shinozaki}, {Shu}, {Stompor}, {Sugai}, {Sugita}, {Suzuki}, {Suzuki},
  {Tajima}, {Takada}, {Takakura}, {Takano}, {Takei}, {Tomaru}, {Tomita},
  {Turin}, {Utsunomiya}, {Uzawa}, {Wada}, {Watanabe}, {Westbrook}, {Whitehorn},
  {Yamada}, {Yamasaki}, {Yamashita}, {Yoshida}, {Yoshida}, \&
  {Yotsumoto}}]{2016JLTP..184..824M}
{Matsumura}, T., {Akiba}, Y., {Arnold}, K., {et~al.} 2016, Journal of Low
  Temperature Physics, 184, 824

\bibitem[{{Meyers} {et~al.}(2017){Meyers}, {Hurley-Walker}, {Hancock},
  {Franzen}, {Carretti}, {Staveley-Smith}, {Gaensler}, {Haverkorn}, \&
  {Poppi}}]{2017PASA...34...13M}
{Meyers}, B.~W., {Hurley-Walker}, N., {Hancock}, P.~J., {et~al.} 2017, \pasa,
  34, e013

\bibitem[{{Miville-Desch{\^e}nes} {et~al.}(2008){Miville-Desch{\^e}nes},
  {Ysard}, {Lavabre}, {Ponthieu}, {Mac{\'{\i}}as-P{\'e}rez}, {Aumont}, \&
  {Bernard}}]{2008A&A...490.1093M}
{Miville-Desch{\^e}nes}, M.-A., {Ysard}, N., {Lavabre}, A., {et~al.} 2008,
  \aap, 490, 1093

\bibitem[{{Pearson} \& {C-BASS Collaboration}(2016)}]{2016AAS...22830104P}
{Pearson}, T.~J. \& {C-BASS Collaboration}. 2016, in American Astronomical
  Society Meeting Abstracts, Vol. 228, American Astronomical Society Meeting
  Abstracts, 301.04

\bibitem[{{Planck Collaboration I}(2016)}]{planck2015-I}
{Planck Collaboration I}. 2016, \aap, 594, A1

\bibitem[{{Planck Collaboration II}(2015)}]{planck2015-II}
{Planck Collaboration II}. 2015, A\&A, submitted, arXiv:1502.01583

\bibitem[{{Planck Collaboration Int. L}(2017)}]{planck2017-L}
{Planck Collaboration Int. L}. 2017, \aap, 599, A51

\bibitem[{{Planck Collaboration Int. LIV}(2018)}]{planck2017-LIV}
{Planck Collaboration Int. LIV}. 2018, ArXiv e-prints: 1801.04945

\bibitem[{{Planck Collaboration Int. XXII}(2015)}]{planck2014-XXII}
{Planck Collaboration Int. XXII}. 2015, A\&A, 576, A107

\bibitem[{{Planck Collaboration Int. XXX}(2016)}]{planck2014-XXX}
{Planck Collaboration Int. XXX}. 2016, \aap, 586, A133

\bibitem[{{Planck Collaboration Int. XXXII}(2016)}]{planck2016-XXXII}
{Planck Collaboration Int. XXXII}. 2016, \aap, 586, A135

\bibitem[{{Planck Collaboration IX}(2016)}]{planck2015-IX}
{Planck Collaboration IX}. 2016, \aap, 594, A9

\bibitem[{{Planck Collaboration X}(2016)}]{planck2015-X}
{Planck Collaboration X}. 2016, \aap, 594, A10

\bibitem[{{Planck Collaboration XII}(2016)}]{planck2015-XII}
{Planck Collaboration XII}. 2016, \aap, 594, A12

\bibitem[{{Planck Collaboration XIII}(2016)}]{planck2015-XIII}
{Planck Collaboration XIII}. 2016, \aap, 594, A13

\bibitem[{{Rubi{\~n}o-Mart{\'{\i}}n} {et~al.}(2017){Rubi{\~n}o-Mart{\'{\i}}n},
  {G{\'e}nova-Santos}, {Rebolo}, {Aguiar}, {G{\'o}mez-Re{\~n}asco},
  {Guti{\'e}rrez}, {Hoyland}, {L{\'o}pez-Caraballo}, {Pel{\'a}ez-Santos},
  {P{\'e}rez-de-Taoro}, {Poidevin}, {Ruiz-Granados}, {S{\'a}nchez de la Rosa},
  {Tramonte}, {Vega-Moreno}, {Viera-Curbelo}, {Vignaga},
  {Mart{\'{\i}}nez-Gonz{\'a}lez}, {Barreiro}, {Casaponsa}, {Casas}, {Diego},
  {Fern{\'a}ndez-Cobos}, {Herranz}, {L{\'o}pez-Caniego}, {Ortiz}, {Vielva},
  {Artal}, {Aja}, {Cagigas}, {Cano}, {de la Fuente}, {Mediavilla}, {Ter{\'a}n},
  {Villa}, {Piccirillo}, {Dickinson}, {Grainge}, {Harper}, {Maffei},
  {McCulloch}, {Melhuish}, {Pisano}, {Watson}, {Lasenby}, {Ashdown}, {Perrott},
  {Razavi-Ghods}, {Titterington}, \& {Scott}}]{2017hsa9.conf...99R}
{Rubi{\~n}o-Mart{\'{\i}}n}, J.~A., {G{\'e}nova-Santos}, R., {Rebolo}, R.,
  {et~al.} 2017, in Highlights on Spanish Astrophysics IX, ed. S.~{Arribas},
  A.~{Alonso-Herrero}, F.~{Figueras}, C.~{Hern{\'a}ndez-Monteagudo},
  A.~{S{\'a}nchez-Lavega}, \& S.~{P{\'e}rez-Hoyos}, 99--107

\bibitem[{{Sheehy} \& {Slosar}(2017)}]{2017arXiv170909729S}
{Sheehy}, C. \& {Slosar}, A. 2017, ArXiv e-prints: 1709.09729

\bibitem[{{Stompor} {et~al.}(2016){Stompor}, {Errard}, \&
  {Poletti}}]{2016PhRvD..94h3526S}
{Stompor}, R., {Errard}, J., \& {Poletti}, D. 2016, \prd, 94, 083526

\bibitem[{{Takayuki Matsuda} \& {the Polarbear
  Collaboration}(2017)}]{2017AAS...23030401T}
{Takayuki Matsuda}, F. \& {the Polarbear Collaboration}. 2017, in American
  Astronomical Society Meeting Abstracts, Vol. 230, 304.01

\bibitem[{{The Polarbear Collaboration}(2014)}]{PolarBear14}
{The Polarbear Collaboration}. 2014, ApJ, 794, 171

\bibitem[{{The Polarbear Collaboration}(2017)}]{PolarBear17}
{The Polarbear Collaboration}. 2017, \apj, 848, 121

\bibitem[{{Thorne} {et~al.}(2017){Thorne}, {Dunkley}, {Alonso}, \&
  {N{\ae}ss}}]{2017MNRAS.469.2821T}
{Thorne}, B., {Dunkley}, J., {Alonso}, D., \& {N{\ae}ss}, S. 2017, \mnras, 469,
  2821

\bibitem[{{Tristram} {et~al.}(2005){Tristram}, {Mac{\'{\i}}as-P{\'e}rez},
  {Renault}, \& {Santos}}]{Tristram05}
{Tristram}, M., {Mac{\'{\i}}as-P{\'e}rez}, J.~F., {Renault}, C., \& {Santos},
  D. 2005, \mnras, 358, 833

\bibitem[{{Zaldarriaga} \& {Seljak}(1997)}]{1997PhRvD..55.1830Z}
{Zaldarriaga}, M. \& {Seljak}, U. 1997, \prd, 55, 1830

\bibitem[{{Zaldarriaga} \& {Seljak}(1998)}]{1998PhRvD..58b3003Z}
{Zaldarriaga}, M. \& {Seljak}, U. 1998, \prd, 58, 023003

\bibitem[{{Zaroubi} {et~al.}(2015){Zaroubi}, {Jeli{\'c}}, {de Bruyn},
  {Boulanger}, {Bracco}, {Kooistra}, {Alves}, {Brentjens}, {Ferri{\`e}re},
  {Ghosh}, {Koopmans}, {Levrier}, {Miville-Desch{\^e}nes}, {Montier}, {Pandey},
  \& {Soler}}]{2015MNRAS.454L..46Z}
{Zaroubi}, S., {Jeli{\'c}}, V., {de Bruyn}, A.~G., {et~al.} 2015, \mnras, 454,
  L46

\end{thebibliography}

\end{document}